\documentclass[aps,reprint,superscriptaddress,amsmath,amssymb,prd,nofootinbib]{revtex4-2}

\usepackage{graphicx}
\usepackage{dcolumn}
\usepackage{bm}

\usepackage{braket} 
\usepackage{color}
\usepackage{amsmath, amssymb}
\usepackage{mathrsfs} 
\usepackage{qcircuit}

\newcommand{\del}{\partial}
\newcommand{\ii}{{\rm i}}

\newcommand{\delsla}{\partial\hspace{-2mm}/\hspace{0.5mm}}

\begin{document}
\title{Quantum data learning for quantum simulations in high-energy physics}

\author{Lento Nagano}
\email{lento@icepp.s.u-tokyo.ac.jp}
\affiliation{International Center for Elementary Particle Physics (ICEPP), The University of Tokyo, 7-3-1 Hongo, Bunkyo-ku, Tokyo 113-0033, Japan}
\author{Alexander Miessen}
\email{lex@zurich.ibm.com}
\affiliation{IBM Quantum, IBM Research -- Zurich, 8803 Rüschlikon, Switzerland}
\affiliation{Institute for Computational Science, University of Zurich, Winterthurerstrasse 190, 8057 Zurich, Switzerland}
\author{Tamiya Onodera}
\email{tonodera@jp.ibm.com}
\affiliation{IBM Quantum, IBM Research-Tokyo, 19-21 Nihonbashi Hakozaki-cho, Chuo-ku, Tokyo, 103-8510, Japan}
\author{Ivano Tavernelli}
\email{ita@zurich.ibm.com}
\affiliation{IBM Quantum, IBM Research -- Zurich, 8803 Rüschlikon, Switzerland}
\author{Francesco Tacchino}
\email{fta@zurich.ibm.com}
\affiliation{IBM Quantum, IBM Research -- Zurich, 8803 Rüschlikon, Switzerland}
\author{Koji Terashi}
\email{terashi@icepp.s.u-tokyo.ac.jp}
\affiliation{International Center for Elementary Particle Physics (ICEPP), The University of Tokyo, 7-3-1 Hongo, Bunkyo-ku, Tokyo 113-0033, Japan}

\date{\today}

\begin{abstract}
Quantum machine learning with parametrised quantum circuits has attracted significant attention over the past years as an early application for the era of noisy quantum processors. 
However, the possibility of achieving concrete advantages over classical counterparts in practical learning tasks is yet to be demonstrated. A promising avenue to explore potential advantages is the learning of data generated by quantum mechanical systems and presented in an inherently quantum mechanical form. 
In this article, we explore the applicability of quantum-data learning to practical problems in high-energy physics, aiming to identify domain specific use-cases 
where quantum models can be employed. We consider quantum states governed by one-dimensional lattice gauge theories and a phenomenological 
quantum field theory in particle physics, generated by digital quantum simulations or variational methods to approximate target states. 
We make use of an ansatz based on quantum convolutional neural networks and numerically show that it is capable of recognizing  
quantum phases of ground states in the Schwinger model, (de)confinement phases from time-evolved states in the~$\mathbb{Z}_2$ gauge theory,
and that it can extract fermion flavor/coupling constants in a quantum simulation of parton shower. 
The observation of non-trivial learning properties demonstrated in these benchmarks will motivate further exploration of the quantum-data learning architecture in high-energy physics.
\end{abstract}

\maketitle


\section{Introduction}
\label{sec:intro}

Quantum computers were originally conceived
to tackle some of the hardest research problems in the physical sciences~\cite{Feynman1982}, such as dynamical simulations of quantum mechanical systems~\cite{RevTacchino2020,Miessen2023}. Recent technological advances brought such a vision closer to reality~\cite{acin_quantum_2018,Altman2021,Alexeev2021,bravyi2022future}.
Among the several application domains that have been associated to noisy, near term quantum computers~\cite{nisqRMP2022}, Quantum Machine Learning (QML)~\cite{Biamonte2017QuantumLearning,Mangini_2021,cerezo2022challenges} received considerable attention in the last years, thanks to its inherently heuristic and versatile nature complemented by some provable performance guarantees~\cite{Liu2021ALearning}. While remarkable progress has been made in understanding the trainability and generalization power (i.e., prediction capability on unseen test data) of QML models based on variational quantum algorithms (VQAs)~\cite{mcclean2018barren,cerezo2021variational,abbas2021power,tacchinoIEEE2021,banchi2021,Caro_Generaliz_2022,LiuPRXQ2022,liu2022}, it remains largely unclear whether, and in which specific domains, QML could complement and surpass classical methods in performing data-driven analysis and learning tasks of concrete practical relevance.

In this context, a particularly promising route is represented by the study of inherently quantum mechanical data, such as quantum states generated in specific experiments or within quantum communication and simulation protocols~\cite{Cong_2019,Uvarov_2020,Lazzarin_2022,Sancho-Lorente_2022,Herrmann_2022,Nathaniel_2021,Wu_2021,Monaco_2022,Caro_Generaliz_2022,2023arXiv230209751N,huang2022quantum,2021arXiv210903400S,Bisio2010,beer2020training,Caro_OutOfDist_2022,Gibbs_2022}.
This approach appears distinctly attractive as it eliminates the necessity of encoding classical data into quantum circuits, leveraging the intrinsic capability of QML architectures to directly manipulate quantum wavefunctions. 
Such QML paradigm that directly exploits quantum forms of data, referred to as quantum-data learning (QDL) in this work, may 
have considerable impact in areas like quantum sensing, quantum many-body physics, quantum chemistry and high-energy/nuclear physics. Paradigmatic QDL problems concern, among others, recognizing quantum phases of matter in strongly correlated systems~\cite{Cong_2019,Uvarov_2020,Lazzarin_2022,Sancho-Lorente_2022,Herrmann_2022,Nathaniel_2021,Wu_2021,Monaco_2022,Caro_Generaliz_2022}, 
classifying model Hamiltonians~\cite{2023arXiv230209751N}, 
predicting expectation values of several incompatible observables on quantum states and performing quantum linear algebra manipulations~\cite{huang2022quantum}, learning entanglement of quantum states~\cite{2021arXiv210903400S}, as well as modeling quantum processes~\cite{huang2022quantum,Bisio2010,beer2020training,Caro_OutOfDist_2022,Gibbs_2022}.

Up to now, most investigations in the field of QDL focused either on abstract scenarios or on conventional models in condensed matter physics such as spin chains.
 In this work, we extend the QDL framework to the realm of high-energy physics, identifying an extensive set of suitable use-cases
and demonstrating -- with numerical calculations -- the direct application of QML methods on some representative problem instances. More concretely, we begin from quantum phase recognition tasks in lattice gauge theory (LGT) for the Schwinger model, which we successfully address with Quantum Convolutional Neural Networks (QCNNs) acting on variationally-generated quantum states. We then move beyond the study of ground state properties by tackling the analysis of dynamical processes, specifically the classification of (de)confinement phases by using time-evolved states for the~$(1+1)$-dimensional~$\mathbb{Z}_2$ gauge theory. In this case, quantum data are generated through a digital quantum simulation using the standard Suzuki-Trotter decomposition. Finally, we consider multi-particle states appearing in parton showering, which can be described by phenomenological quantum field theory (QFT) models. Here, we employ QDL to classify fermion flavors and to predict coupling constants from quantum states generated by a quantum circuit implementation of such phenomenological QFT simulations, a task akin to Hamiltonian learning~\cite{Bairey2019,LiPRL2020}. Overall, our results suggest that QDL methods, and particularly QCNNs, can become useful resources for current and future quantum-powered investigations in high-energy physics, both on the computational side (e.g., when working with quantum simulators) and in close connection with experiments (e.g., when employing quantum devices).

The paper is organized as follows. In Sec.~\ref{sec:methods}, we start from a generic quantum circuit model for QML and we illustrate the specific QDL architecture using QCNNs employed in this work. In Sec.~\ref{sec:schwinger}, we investigate the task of quantum phase recognition in the Schwinger model by applying QDL methods to variationally-generated quantum data and give the numerical results. Sec.~\ref{sec:z2} discusses the classification of confined and deconfined phases of the $\mathbb{Z}_2$ gauge theory. In Sec.~\ref{sec:qps}, we describe the QDL method employed for determining coupling constants and fermion flavors from multi-particle states in parton showering. Finally, we conclude in Sec.~\ref{sec:conclusion} with a discussion on the results and an outlook on possible future research directions.

\section{Methods}
\label{sec:methods}

\begin{figure*}[t!]
    \centering
    \includegraphics[width=0.7\linewidth]{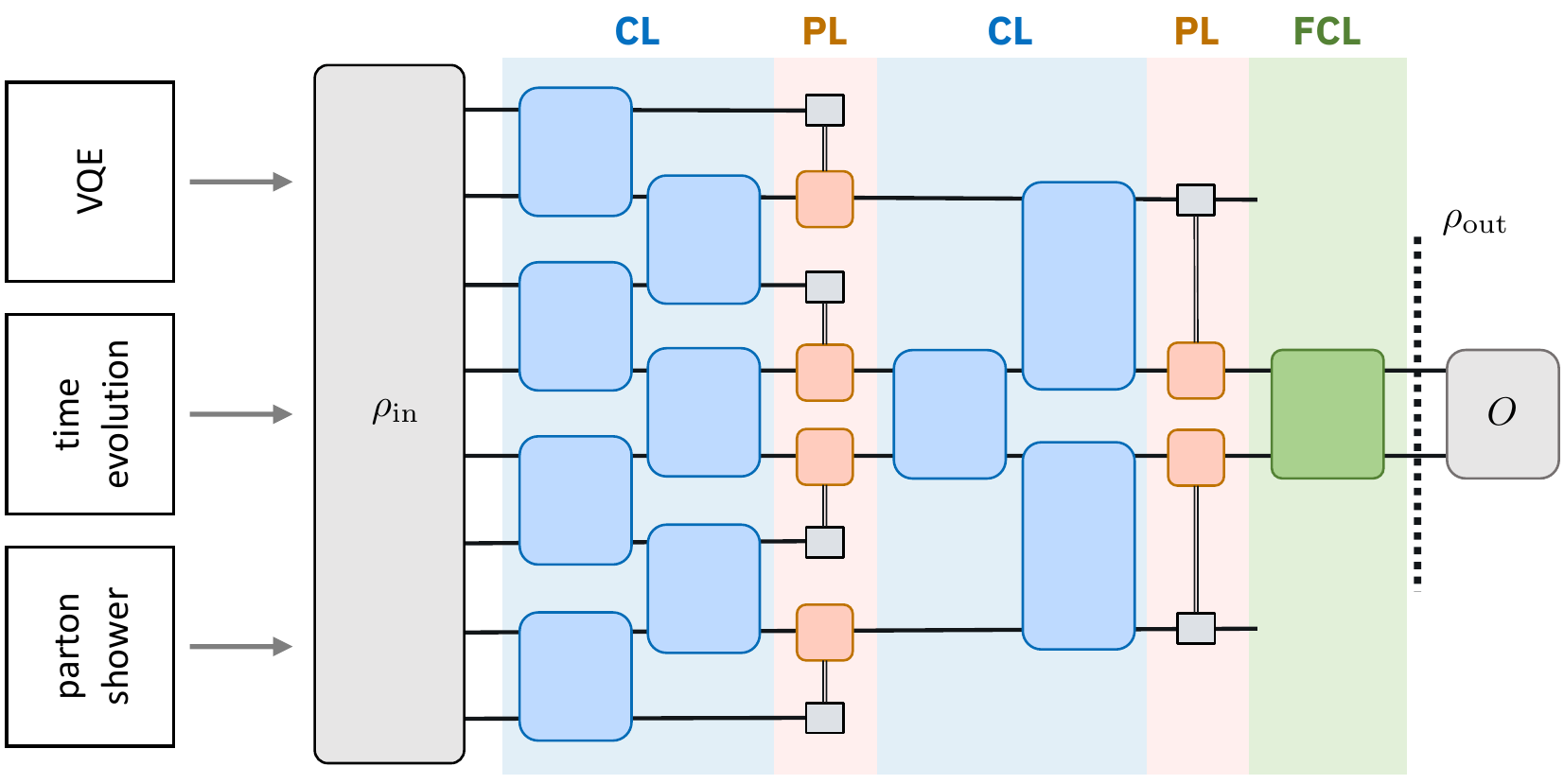}
    \caption{Representative QCNN circuit considered in this work, composed of alternating convolution and pooling layers (denoted by CL and PL, respectively), followed by a fully connecter layer (FCL) and the measurement of the output state $\rho_{\text{out}}$ with an observable $O$. In the QDL framework used in this paper, the input to the QCNN circuit is a quantum state $\rho_{\text{in}}$, either in the form of a ground state generated with VQE, a time-evolved state through Hamiltonian simulation, or a multi-particle state from a phenomenological quantum parton shower simulation. Precise QCNN circuits used in the numerical studies are described in Sec.~\ref{sec:schwinger}, \ref{sec:z2}, and \ref{sec:qps}, with details in Appendix~\ref{app:qcnn}.}
    \label{fig:QCNN-circ}
\end{figure*}

We adopt the approach of supervised QML with a parameterized quantum circuit, often referred to as Quantum Neural Network (QNN)~\cite{benedetti2019parameterized}.
In the QDL framework, the inputs to the parameterized circuit are represented by quantum states prepared on a quantum processor.
In the applications presented in this work, input quantum states may implicitly encode some classical parameters~${\boldsymbol x}$, such as certain values of coupling constants appearing in an underlying Hamiltonian.
We generically write~$\rho_{\rm in}\equiv\rho(\boldsymbol x)$ for input states chosen among a training dataset~$\{(\rho(\boldsymbol x), y_{\boldsymbol x})\}_{\bm{x}\in\mathcal{T}_{\text{train}}}$.
The labels~$y_{\boldsymbol x}$ denote, for example, certain physical phases, properties of the quantum state, or Hamiltonian parameters that the model should learn to recognize or predict.
In practice, for any given~${\boldsymbol x}$, the corresponding input state~$\rho(\boldsymbol x)$ will be prepared by a unitary~$U_{\rm prep}(\boldsymbol x)$.
Depending on the problem,~$U_{\rm prep}(\boldsymbol x)$ can take various forms as depicted in Fig.~\ref{fig:QCNN-circ}, including a VQE ansatz~\cite{peruzzo2014variational} approximating the ground state of a Hamiltonian~$H(\boldsymbol x)$, a circuit implementing a time-evolution~$U=e^{-iH(\boldsymbol x)t}$~\cite{RevTacchino2020}, or another phenomenological quantum simulation circuit. Each input state is processed by a QNN circuit~$U_{\rm QNN}({\boldsymbol \theta})$ with trainable parameters~${\boldsymbol \theta}$, producing the state~$\rho_{\rm out}(\boldsymbol x,{\boldsymbol \theta})=U_{\rm QNN}({\boldsymbol \theta})\rho(\boldsymbol x) U_{\rm QNN}^\dagger({\boldsymbol \theta})$.
This state is then measured to retrieve a model output~$y_{\rm out}(\boldsymbol x, \boldsymbol \theta)$ as the expectation value~$\langle O\rangle={\rm Tr}[\rho_{\rm out}O]$ of a suitable Hermitian operator~$O$.
The learning is carried out by optimizing the set of parameters ${\boldsymbol \theta}$, with the aim of minimizing a cost function ${\cal L}(\bm{\theta})$ that quantifies the distance of the model predictions $\{y_{\rm out}(\boldsymbol x, \boldsymbol \theta)\}_{\boldsymbol x\in\mathcal{T}_{\text{train}}}$ from the ideal training labels~${\{y_{\boldsymbol x}\}_{\boldsymbol x\in\mathcal{T}_{\text{train}}}}$.

As a QNN ansatz, we employ the class of models known as QCNNs~\cite{Cong_2019}. QCNNs are closely related to the multiscale entanglement renormalization ansatz (MERA) and are known to be particularly well suited for paradigmatic QDL applications such as quantum phase recognition~\cite{Cong_2019,Lazzarin_2022,Herrmann_2022,Nathaniel_2021,Monaco_2022}.
Specifically, this class of models is expected to achieve advantages in terms of sampling complexity when conventional methods rely on non-local order parameters~\cite{Cong_2019}. Our choice is further motivated by the successful application of Convolutional Neural Networks (CNNs) to similar tasks~\cite{Carrasquilla_2017} in classical machine learning.
The representative QCNN circuit $U_{\text{QCNN}}(\bm{\theta})$, which is schematically shown in Fig.~\ref{fig:QCNN-circ}, consists of alternating convolutional and pooling layers: the former apply translationally-invariant unitary gates to local subsets of qubits while the latter reduce the dimensionality of the state by local measurements and classical feed forward.
At the final step, a fully connected layer is applied globally to the remaining qubits, followed by the measurement of an observable~\footnote{We give the precise form of the QCNN circuit used in this study in Appendix~\ref{app:qcnn}. As explained there, the translational invariance of the QCNN parameters is not maintained in the study with the $\mathbb{Z}_2$ gauge theory simulation.}.
Thanks to their naturally shallow circuit structure, QCNNs are provably resilient to the phenomenon of barren plateaus, which hinders the trainability of more generic QNN models~\cite{Pesah2021}.

With a QCNN model $U_{\text{QCNN}}$, the supervised QDL architecture introduced above can be further detailed as follows. For $N$-qubit input quantum data of the form $\rho(\boldsymbol x) = |\psi_{\boldsymbol x}\rangle\langle \psi_{\boldsymbol x}|$, we construct the model output~$y_{\rm out}(\boldsymbol x, \boldsymbol \theta)$ by measuring the last qubit in the $Z$ basis, namely
\begin{equation}
    y_{\rm out}(\boldsymbol x, \boldsymbol \theta)=\Braket{\psi_{\boldsymbol x} |U_{\text{QCNN}}^{\dag}cZ_{N-1}U_{\text{QCNN}}|\psi_{\boldsymbol x}}\, ,
\end{equation}
where $c=1$ in the case of the Schwinger model (Sec.~\ref{sec:schwinger}) and of the $\mathbb{Z}_2$ gauge theory (Sec.~\ref{sec:z2}).
In the case of the parton shower (Sec.~\ref{sec:qps}), we set $c=2$  to increase the output range in the regression of the coupling constants. The output is then used by a classical optimizer that solves the minimization problem:
\begin{equation}
\bm{\theta}_{\text{opt}}=\arg\min_{\bm{\theta}}\mathcal{L}(\bm{\theta})\,.
\end{equation}
The cost function is 
\begin{equation}
\mathcal{L}(\bm{\theta})=\frac{1}{\left|\mathcal{T}_{\text{train}}\right|}\sum_{\boldsymbol x\in\mathcal{T}_{\text{train}}}(y_{\boldsymbol x}-y_{\rm out}(\boldsymbol x, \boldsymbol \theta))^{2}\,
\end{equation}
for the tasks with the Schwinger model, $\mathbb{Z}_2$ gauge theory and the coupling constant determination in the parton shower, and
\begin{align}
\mathcal{L}(\bm{\theta}) = \, & - \frac{1}{\left|\mathcal{T}_{\text{train}}\right|}\sum_{\boldsymbol x\in\mathcal{T}_{\text{train}}} \Bigl[ y_{\boldsymbol x} \log \bigl(y_{\rm out}(\boldsymbol x, \boldsymbol \theta) \bigr) \nonumber \\
& + (1-y_{\boldsymbol x})\log \bigl(1-y_{\rm out}(\boldsymbol x, \boldsymbol \theta) \bigr) \Bigr] \,
\end{align}
for the 2-class flavor classification task in the parton shower, where $\left|\mathcal{T}_{\text{train}}\right|$ is the number of training data.
With the optimized parameters $\bm{\theta}_{\text{opt}}$, the trained QCNN model provides $y_{\rm out}(\boldsymbol x, \boldsymbol \theta_{\text{opt}})$, the prediction label for the test data $\boldsymbol x\in\mathcal{T}_{\text{test}}$.
The number of trainable parameters~${\boldsymbol \theta}$ used in the studies is $21N_L$ with $N_L$ being the number of pairings of convolution and pooling layers (see Appendix~\ref{app:qcnn} for the forms of parameterized unitary gates).
All results presented in this work are obtained from noiseless numerical simulations, manipulating exact representations of statevectors and unitary operations.

\section{Quantum simulation of Schwinger model}
\label{sec:schwinger}

In this section, we address a phase recognition problem for the Schwinger model~\cite{PhysRev.128.2425}, a $(1+1)$-dimensional U(1) LGT (see also e.g.~\cite{Martinez:2016yna,Klco:2018kyo} for digital quantum simulation of this model).
The aim is to learn quantum states provided as input data and their associated phases as labels and predict phases of unknown states in test data.
We first introduce the Schwinger model and its phases, then describe how we prepare the input dataset using VQE and construct the QCNN model. Finally we present some numerical results.

\subsection{Schwinger model}
\label{subsec:schwinger-model}

The Lagrangian of the continuum Schwinger model is given by
\begin{align}
\mathscr{L}_{\rm Schwinger}
= & -\frac{1}{4} F_{\mu\nu} F^{\mu\nu} +\ii\bar{\psi}\gamma^\mu (\del_\mu +\ii gA_\mu -m) \psi
\nonumber\\
&
+\frac{g\vartheta}{4\pi} \epsilon_{\mu\nu} F^{\mu\nu}
\,,
\end{align}
where the first term is the kinetic term for a gauge field~$A_{\mu}$, the second term comprises the kinetic and mass term of a Dirac fermion~$\psi$ as well as its coupling to the gauge field, and the last term is a topological term.
We can obtain a lattice Hamiltonian via the staggered fermion formalism~\cite{Kogut:1974ag} where fermions are defined on the lattice sites and gauge fields are defined on the links between adjacent sites.
In $1$D, we can further integrate out all gauge fields making use of the open boundary conditions and Gauss's law.
Then, applying the Jordan-Wigner transformation~\cite{Jordan1928} to map fermionic to qubit operators, the Hamiltonian is written in terms of Pauli matrices as
\begin{align}
 H = \, & J\sum_{n=0}^{N_s-2} \left[\sum_{i=0}^{n}\frac{Z_i + (-1)^i}{2}
 +\frac{\vartheta}{2\pi}
 \right]^2 
 \notag \\
 & + \frac{w}{2}\sum_{n=0}^{N_s-2}\big[X_n X_{n+1}+Y_{n}Y_{n+1}\big]
 + \frac{m}{2}\sum_{n=0}^{N_s-1}(-1)^n Z_n
 \, ,
 \label{eq:hamiltonian-schwinger}
\end{align}
where $J=ag^{2}/2$, $w=1/(2a)$, and $N_s$ is the number of spatial lattice sites.

The continuum model is known to exhibit a phase transition
at~$\vartheta=\pi$ and~$(m/g)=(m/g)_{c}\approx 0.33$ due to breaking of the {CT} symmetry~\cite{Coleman:1976uz,Byrnes:2002gj,PhysRevD.95.094509}.
A simple order parameter characterizing this phase transition is the expectation value of an averaged electric field:
\begin{equation}
E=\frac{1}{N}\sum_{n=0}^{N_s-1}\sum_{i=0}^{n}\frac{Z_i + (-1)^i}{2}\, .
\label{eq:Schwinger-E-field}
\end{equation}
For $\vartheta=\pi$, we have $\braket{E}=0$ below the critical mass (symmetric phase) and $\braket{E}\neq0$ above the critical mass (symmetry breaking phase).

\subsection{Simulation method}
\label{subsec:method-schwinger-qcnn}
In this first application, we employ VQE
to prepare an input dataset $\{\ket{\psi_m}, y_m\}_{m\in\mathcal{M}}$.
Each input state~$\ket{\psi_m}$ is the ground state corresponding to the Hamiltonian~$H$ in Eq.~\eqref{eq:hamiltonian-schwinger} at given mass $m \in \mathcal{M}$, with $y_m=\pm1$ being the respective label characterizing the phase:
\begin{equation}
y_{m}=
\begin{cases}
+1 & (m/g)>(m/g)_c
\\
-1 & (m/g)<(m/g)_c
\end{cases}\,.
\end{equation}
We split the full data into training and test data as~$\{\ket{\psi_m}, y_m\}_{m\in\mathcal{M}_\text{train/test}}$ with $\mathcal{M}=\mathcal{M}_\text{train}\sqcup\mathcal{M}_\text{train}$ being a disjoint union of the training and test data.
Our goal is to train the QCNN with the training data and predict labels associated with the test data.

We prepare the ground states by performing VQE using the Hamiltonian variational ansatz (HVA)~\cite{PhysRevA.92.042303, 10.21468/SciPostPhys.6.3.029, PRXQuantum.1.020319}~$\ket{\psi(\bm{\lambda})}=U_{\text{HVA}}(\bm{\lambda})\ket{\phi}$, with variational parameters~$\bm{\lambda}$.
The precise form of the ansatz is given in Appendix~\ref{sec:HVA-schwinger}.

The initial state is fixed as~$\ket{\phi} = \ket{01}^{N_s/2}$ to constrain the minimization to the zero-magnetization sector,~$\sum_i Z_i \ket{\psi}= 0$.
The optimal parameters are then found via classical optimization of the Hamiltonian expectation value:
\begin{equation}
\bm{\lambda}_{\text{opt}}(m)
=\arg\min_{\bm{\lambda}}\braket{\psi(\bm{\lambda})|H(m)|\psi(\bm{\lambda})}\,,
\end{equation}
where we explicitly write $H(m)$ to indicate that the mass is a parameter of the Hamiltonian.
The input data is then produced by $\ket{\psi_{m}}:=\ket{\psi(\bm{\lambda}_{\text{opt}}(m))}$.

\subsection{Results}
\label{subsec:results-schwinger}
Before discussing the results in detail, we briefly summarize the simulation setup.
We obtain ground states of the Schwinger model with $N=N_s=8, ag=2, \vartheta=\pi$, $m/g\in\mathcal{M}=\{-2+0.05\text{n} \, | \, \text{n}=0,1,\ldots,80\}$.
The critical mass takes the value $(m/g)_c\approx0.143$ for $ag=2$ as obtained from the exact diagonalization (ED) (see Appendix~\ref{subsec:critical-mass-schwinger} for the derivation).
The approximate ground states were found using VQE with SLSQP as the classical optimizer.
For better convergence, we initialized the variational parameters with the optimal parameters found using the previous $m/g$ value.
Even with this precaution in place, VQE performance tends to deteriorate during and after crossing the phase transition: therefore, we also scanned over the $\mathcal{M}$ from both directions (i.e., one scan starting from $-2$ and the other from 2).
The final input data was subsequently obtained by choosing the optimized parameters corresponding to the smallest energy value found from both scans for a given $m/g$.
As for the QCNN training, the pairing of the convolution and pooling layers is repeated three times ($N_L=3$), and the COBYLA optimizer is used for the classical optimization with $200$ iterations.
This training process is repeated 20 times starting from different random initial parameters.

\begin{figure}[t!]
  \centering
   \includegraphics[width=0.96\columnwidth]{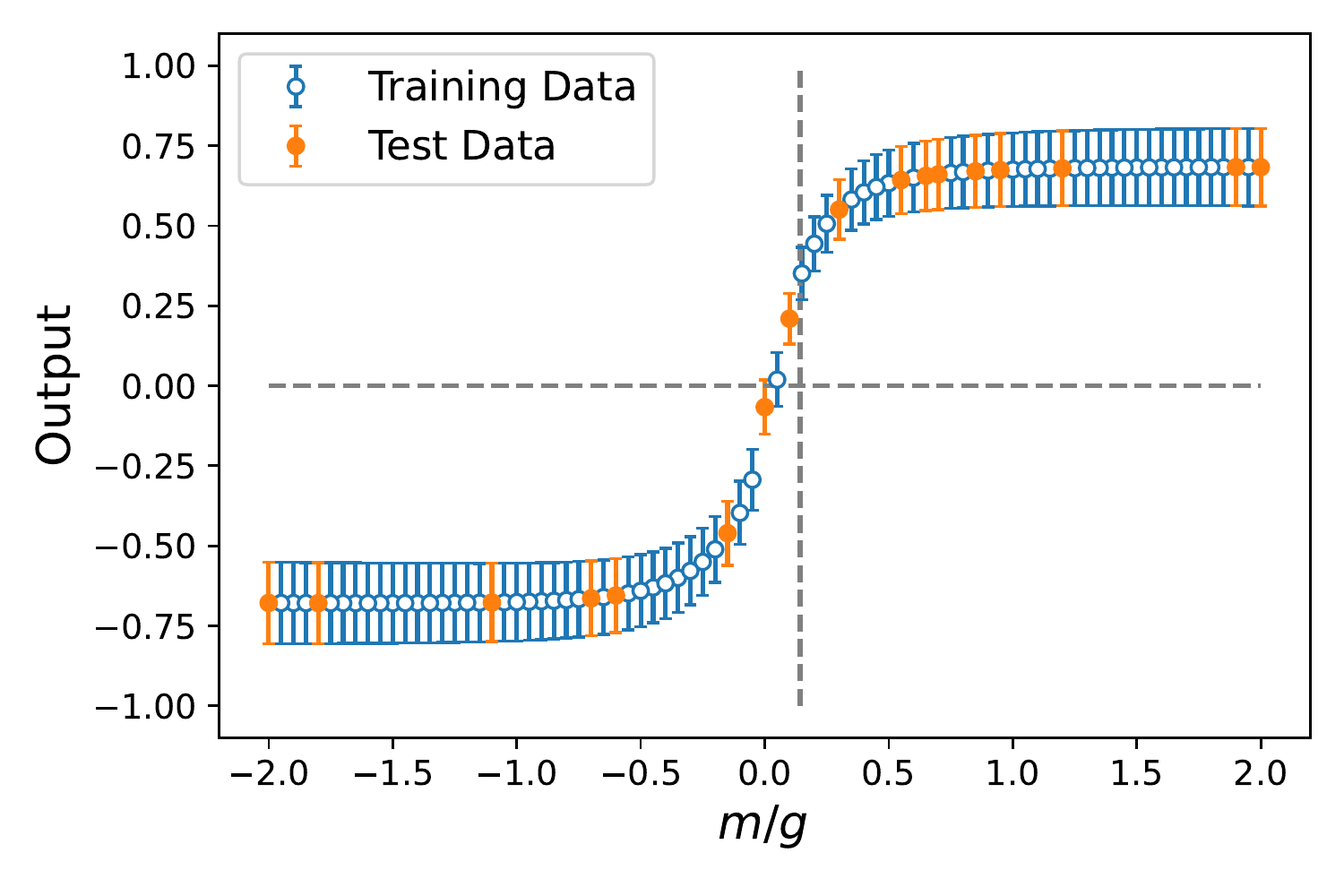}
  \caption{Outputs from the QCNN 
  after the training for the Schwinger model with $N_s=8$, $ag=2$, $\vartheta=\pi$. The blue open (orange filled) markers represent the average of~$20$ trials for training/test data. The bars represent the standard deviations.
  The dashed lines indicate the critical mass value~$(m/g)_c$ obtained from ED (vertical) and~$y_m=0$ (horizontal), respectively.}
  \label{fig:qcnn-schwinger-phase-transition-n8}
\end{figure}

The outputs ${y}_{\text{out}}(m,\bm{\theta})$ from the QCNN circuit after training are shown in Fig.~\ref{fig:qcnn-schwinger-phase-transition-n8}.
Except for a few points, one can correctly distinguish the labels of both training
and test data using the sign of the QCNN output, implying a good prediction ability of the QCNN.

It is important to mention here that, while these proof-of-principle experiments demonstrate the applicability of QDL methods to a paradigmatic learning task in high-energy physics -- and particularly the ability to recognize structure in the model phase space without explicitly introducing an order parameter --, a curve similar to the one appearing in Fig.~\ref{fig:qcnn-schwinger-phase-transition-n8} can be easily reproduced by directly measuring the electric field as given in Eq.~\eqref{eq:Schwinger-E-field}, which would require only single-qubit measurements (see also Appendix~\ref{subsec:local-observable-schwinger}).

\section{Real-time evolution of $(1+1)$d $\mathbb{Z}_2$ gauge theory}
\label{sec:z2}
\subsection{$(1+1)$D $\mathbb{Z}_2$ gauge theory}
In this section, we consider a one-dimensional $\mathbb{Z}_{2}$ gauge theory with staggered fermionic matter~\cite{Kogut:1974ag,Mildenberger:2022jqr,2023arXiv230502361C,2023arXiv230315519B}.
As with the Schwinger model, fermions are defined on lattice sites while gauge fields are defined on the links between sites.
The Hamiltonian of the $\mathbb{Z}_{2}$ gauge theory on a $1$D lattice with $N_s$ sites is given by
\begin{align}
H=&-\frac{J}{2}\sum_{n=0}^{N_s-1}({X}_{n}{Z}_{n,n+1}{X}_{n+1}+{Y}_{n}{Z}_{n,n+1}{Y}_{n+1})
\notag\\
&-f\sum_{n=0}^{N_s-2}{X}_{n,n+1}+\frac{m}{2}\sum_{n=0}^{N_s-1}(-1)^{n}{Z}_{n}\,,
\end{align}
where the Pauli matrix ${P}_n$ with $P\in\{X,Y,Z\}$ acts on site~$n$, and the ${P}_{n,n+1}$ acts on the link between sites~$n$ and~$(n+1)$.
The first term represents the covariant kinetic contribution of fermions, the second represents the background~$\mathbb{Z}_2$ gauge fields, and the third term is the fermion mass term. We take the periodic boundary condition here, i.e., define $P_{N_s}:=P_{0}$.
Physical states must satisfy the gauge invariant condition:
\begin{align}
G_{n}\ket{\text{phys}} = & g_{n}\ket{\text{phys}}
\,,
\label{eq:gauss-law-z2}
\\
G_{n} := & -{X}_{n-1,n}{Z}_{n}{X}_{n,n+1}\,,
\end{align}
where $g_{n}$ is a constant.
We will 
set~$g_{n_\text{probe}}=+1$ and~$g_{n}=-1$ for~$n\neq n_{\text{probe}}$ to
study the effects of a probe charge at site $n_{\text{probe}}$.

This model is known to exhibit two phases~\cite{2019NatPh..15.1168S}. 
The matter fields are confined when a background field is present~($f\neq0$), meaning that the effects of matter fields do not spread out, while they are deconfined in the absence of the background field~($f=0$).

\subsection{Simulation method}
We consider the classification of time-evolved states according to the value of $f$.
Input data are defined by~$\ket{\psi_{m,f}}=e^{-\ii H(m,f)T} \ket{\psi_{\text{init}}}$ at a fixed time slice $t=T$, where we explicitly write the dependence of Hamiltonian on mass~$m$ and background field $f$.
To obtain the dataset, 
the time evolution is approximated through a Suzuki-Trotter decomposition applied to the initial reference state $\ket{\psi_{\text{init}}}$ satisfying Gauss's law in Eq.~\eqref{eq:gauss-law-z2} with ~$n_\text{probe}=1$
(see {Appendix~\ref{sec:circuit-z2}} for details).
The label $y_{m,f}$ associated with a state $\ket{\psi_{m,f}}$ is given by 
\begin{equation}
y_{m,f}=
\begin{cases}
+1 & f\neq 0 \quad (\text{confine})
\\
-1 & f=0 \quad (\text{deconfine})
\end{cases}\,.
\end{equation}
The resulting dataset $\{\ket{\psi_{m,f}}, y_{m,f}\}_{(m,f)\in\mathcal{M}\times \mathcal{F}}$ is split into training and test data, as in Sec.~\ref{subsec:method-schwinger-qcnn}.
In this study, we used a QCNN circuit whose parameters are not translationally invariant. See Appendix~\ref{app:qcnn} for details.

\begin{figure}[t!]
\centering
\includegraphics[width=0.96\columnwidth]{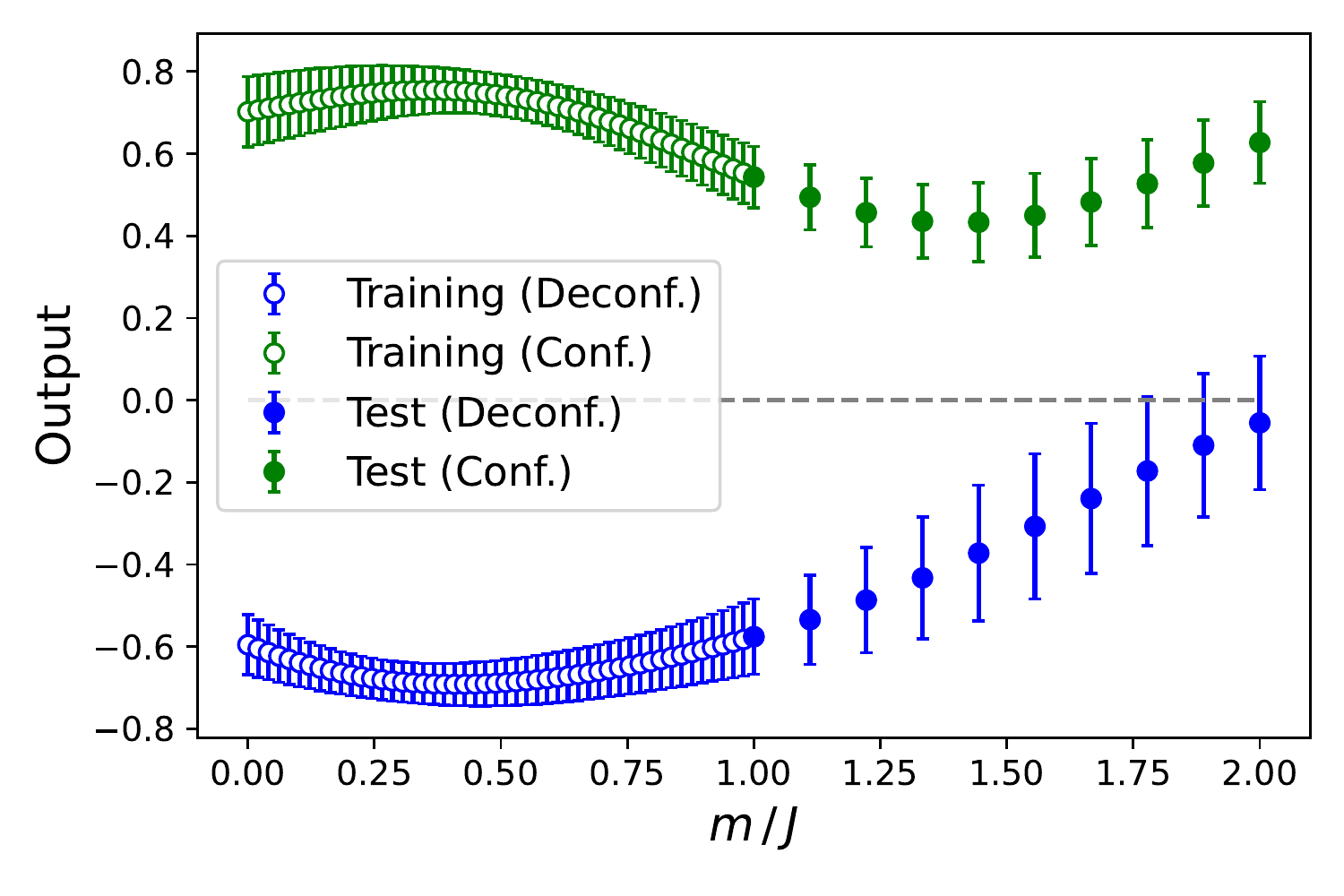}
\caption{Outputs from the QCNN circuit after training for the $\mathbb{Z}_2$ gauge theory with $N_s=2$. The input states are taken at $T=2$. Open (filled) markers represent the average of 20 trials for training (test) data. Bars represent the standard deviations. Green and Blue markers represent confinement ($f/J=3$) and deconfinement ($f/J=0$), respectively.}
\label{fig:z2-qcnn-output}
\end{figure}

\subsection{Results}
We fix the system parameters to $N_{s}=2$, $T=2/J$, and $\mathcal{F}=\{0,3J\}$.
Using periodic boundary conditions, the model is mapped to $N=2N_s=4$ qubits (two for the sites, two for the links).
We prepare the dataset using the Suzuki-Trotter decomposition with~$20$ time steps.
To see the generalization ability to data with larger mass values than those of the training data, we use as the training and test datasets~$\mathcal{M}_\text{train}=\{\text{n}/49 \, | \, \text{n}=0,1,\ldots,49\}$ and~$\mathcal{M}_{\text{test}}=\{1+\text{n}/9 \, | \, \text{n}=0,1,\ldots,9\}$, respectively, and train the QCNN circuit with~$N_L=3$, starting from~$20$ different initializations with~$200$ iterations each.

The QCNN outputs after the training are shown in Fig.~\ref{fig:z2-qcnn-output}.
As one can see from the figure, a clear separation is learned for the two phases of the training dataset. The model also generalizes well for the test data up to~$m\sim1.75$; 
after this value the separation becomes less evident, as it can be expected given the increasing differences between the test and training data. 
It is also worth remarking that, contrary to the Schwinger model example presented in Sec.~\ref{sec:schwinger}, 
there is no known simple local order parameter associated with symmetry breaking by which the two phases can be distinguished.

\section{Quantum parton shower simulation}
\label{sec:qps}

As a third benchmark for the QDL approach, we consider the quantum simulation of multi-particle states governed by a simple QFT model~\cite{Nachman_2021}. The Lagrangian of this model (see Eq.~(1) of~\cite{Nachman_2021}) is
\begin{align}
\mathscr{L}_{\text{PS}}
= \, &\bar{f}_1(i\delsla+m_1)f_1+\bar{f}_2(i\delsla+m_2)f_2+(\del_\mu\phi)^2 \nonumber \\
&+g_1\bar{f}_1f_1\phi+g_2\bar{f}_2f_2\phi+g_{12}[\bar{f}_1f_2+\bar{f}_2f_1]\phi.
\label{eq:QPS_lagr}
\end{align}
The first three terms in Eq.~(\ref{eq:QPS_lagr}) describe the kinematics of the (anti-)fermions $f_i$ ($\bar{f}_i$) and a scalar boson $\phi$, while the last three terms govern their interactions with coupling constants $g_\alpha\, (\alpha=1,2,12)$.
The fermions have two distinct flavors labeled $1$ and $2$, and can radiate the boson ($f_i\to f_j\phi$) with different strengths controlled by the coupling constants:~$g_1$ for~$f_1\to f_1\phi$,~$g_2$ for~$f_2\to f_2\phi$, and~$g_{12}$ for~$f_1\to f_2\phi$ or~$f_2\to f_1\phi$ (and analogously for~$\bar{f}_i$).
In contrast to~\cite{Nachman_2021}, here we do not consider, for simplicity, the channel $\phi\to f_i\bar{f}_j$.
Such radiation and splitting processes are basic building blocks in the simulation of parton showers (PSs) in high-energy physics.
We will therefore refer to the simulation method for this model, explained below, as the quantum PS (QPS) algorithm.

The QPS algorithm starts with an initial fermion~$f_{i(=1,\,2)}$ and repeats the~$f_i\to f_j\phi$ process~$N_{\rm step}$ times to simulate a shower of the fermions and bosons.
In the end, there are at most $N_{\rm step}$ bosons $\phi$.
If $g_{12}$ takes a non-zero value, interference occurs due to different fermion flavors in the intermediate {\it unobserved} states, resulting in different numbers of emissions and kinematic properties of the shower compared to the $g_{12}=0$ case. This phenomenon represents a quantum property of parton showers that is hard to simulate classically~\cite{Nachman_2021}.

\subsection{QPS simulation data}
Two different tasks are considered here, both aiming to predict the parameters in the Lagrangian of Eq.~\eqref{eq:QPS_lagr} by learning the state of a produced $\phi$ boson system.
The first task is to predict the values of the coupling constants by means of regression, and the second is to classify the flavor of the initial fermion.

\begin{figure}[t!]
\centering
\includegraphics[width=0.75\columnwidth]{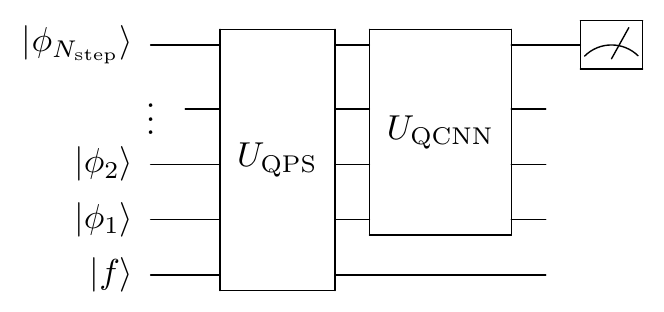}
\includegraphics[width=\columnwidth]{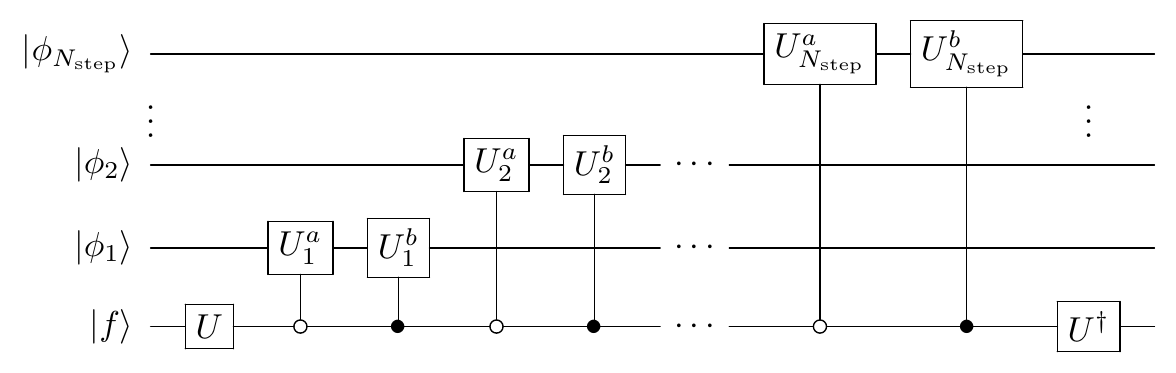}
\caption{(Top) QPS circuit ($U_{\rm QPS}$) to produce the dataset, followed by the QCNN circuit ($U_{\rm QCNN}$) and the single-qubit measurement. (Bottom) Unitary gates comprising the $U_{\rm QPS}$ circuit. The circuit diagram follows the notation used in Ref.~\citenum{Nachman_2021}. The unitaries $U$ and $U_i^{a/b}$ corresponding to Eqs.~(11) and (43) in the supplemental material of Ref.~\citenum{Nachman_2021}, respectively, are given in Appendix~\ref{app:qps}.}
\label{fig:QPS-QCNN-circ}
\end{figure}

Fig.~\ref{fig:QPS-QCNN-circ} shows the circuit used to create the QPS data and to learn the states using QCNN. The QPS circuit ($U_{\text{QPS}}$) to produce the data consists of single-qubit~$U$ gates and two-qubit controlled-$U_i^{a/b}$ gates for $i$-th emission step (see Appendix~\ref{app:qps} for details). The $U_i^{a/b}$ gate  depends implicitly on the emission scale $\theta_i$ through a Sudakov factor that controls the splitting probability of $f_i\to f_j\phi$~\cite{Nachman_2021}.
The $N_{\rm step}$-qubit register that represents the state of emitted $\phi$ bosons $\{\ket{\phi_i}\}_{i=1}^{N_{\text{step}}}$ is fed into the~$U_{\text{QCNN}}$. The $\ket{\phi_i}$ state represents whether a $\phi$ boson is emitted or not at the $i$-th step. 

First, the QPS dataset is produced using the $U_{\text{QPS}}$ with $N_{\rm step}=8$ and an initial fermion selected randomly between $f_1$ and $f_2$. The dataset contains 30 parton showers, 20 of which are used for the training of the $U_{\text{QCNN}}$ and 10 for the testing. In a single experiment, the trainable parameters of the $U_{\text{QCNN}}$ are initialized randomly and are optimized using the 20 training parton showers. In the testing stage, the trained $U_{\rm QCNN}$ model is applied to the 10 testing parton showers. This experiment is repeated 30 times, each starting with random initial QCNN parameters, to obtain the average and the uncertainty of the average, which are reported below. Exactly the same QPS dataset is used in all experiments, and we choose this approach to assess the performance of QCNN model and characterize its network structure by altering the $U_{\rm QCNN}$ into different $U_{\rm QNN}$ models, as discussed below.

\subsection{QCNN for parameter prediction}
The QCNN circuit structure and gates used in the study of QPS are slightly different from those used in the previous two sections (Sec.~\ref{sec:schwinger}-\ref{sec:z2}), as detailed in Appendix~\ref{app:qcnn}.
The number of pairings of the convolution and pooling layers is $N_L=3$ for the QPS dataset with $N_{\rm step}=8$. 
The measurement of $\rho_{\rm out}$ is always performed on the last qubit corresponding to $\ket{\phi_{N_{\rm step}}}$, irrespective of $N_L$, and the observable is chosen to be $2Z$ for both tasks, as motivated in Sec.~\ref{sec:methods}.

For comparison with the QCNN model, an alternative ansatz, referred to as the hardware-efficient ansatz (HEA)~\cite{2017Natur.549..242K}, is considered by replacing $U_{\text{QCNN}}$ in Fig.~\ref{fig:QPS-QCNN-circ} with the corresponding unitary $U_{\text{HEA}}$. The HEA consists of a layer of single-qubit $R_Y$ and $R_Z$ gates on each qubit, followed by two-qubit CZ gates between neighboring qubits with a periodic boundary condition and another layer of the single-qubit rotation gates. The pairing of the layers of single-qubit and two-qubit gates is repeated~$N'_L$ times, leading to the number of trainable parameters of~$2N_{\rm step}(N'_L+1)$. 

In each experiment, the QCNN and HEA circuits are independently set with randomly-chosen initial parameters and trained. 
For the parameter optimization, the COBYLA optimizer is used for both tasks with 2000 iterations. 

\subsection{Results}
\label{subsec:qps_results}
\subsubsection{Prediction of coupling constants}
\begin{figure}[t!]
    \centering
    \includegraphics[width=0.96\columnwidth]{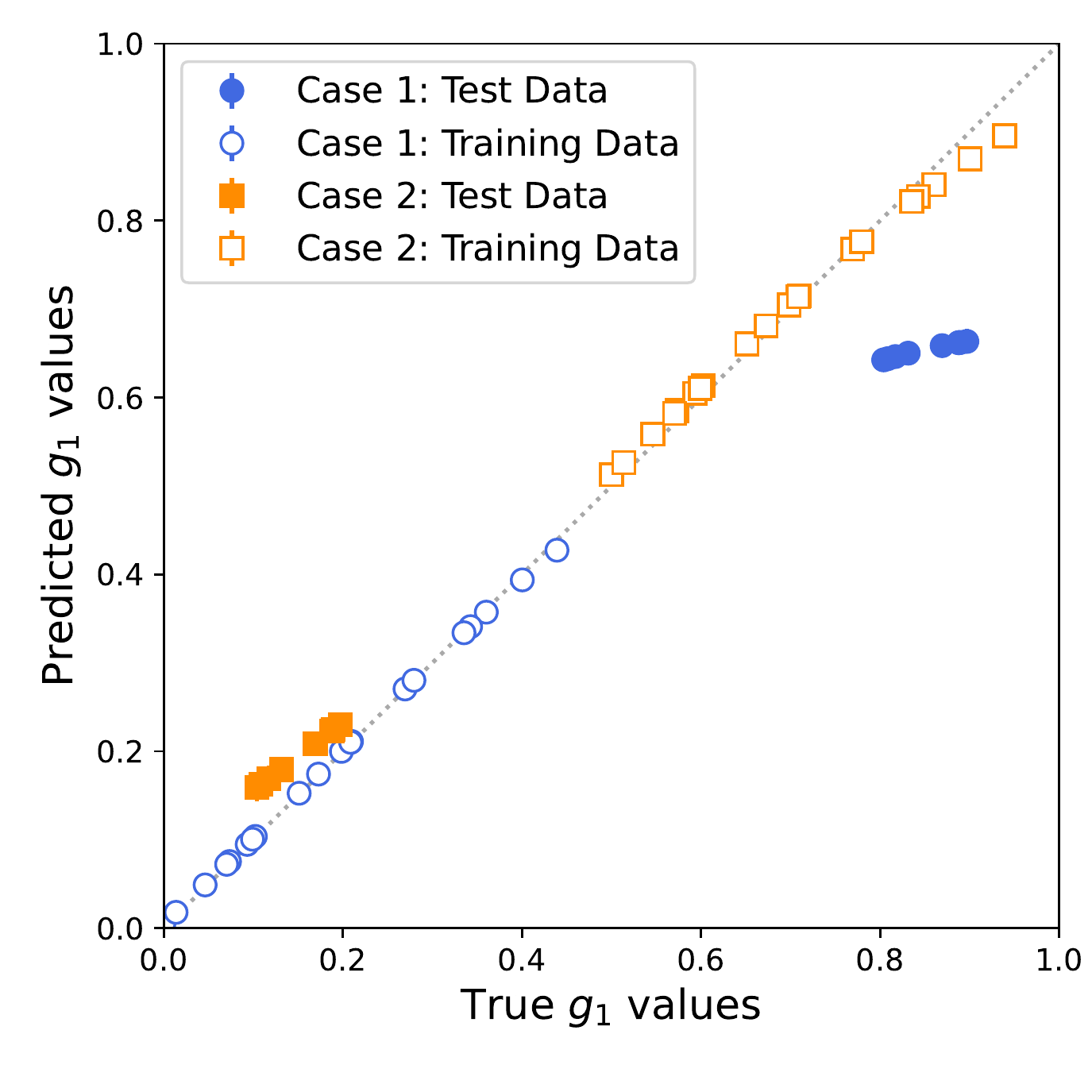}
    \caption{
    Determination of $g_1$ values at $N_{\rm step}=8$. Open (filled) markers represent training (test) data. The circles (squares) correspond to Case 1 (2) of the training data within $0<g_1<0.5$ ($0.5<g_1<1.0$) and the test data within $0.8<g_1<0.9$  ($0.1<g_1<0.2$). The other coupling constants are set to $g_2=1$ and $g_{12}=0$.}
    \label{fig:qps_coupling}
\end{figure}

The prediction of the coupling constants is performed by restricting the training data to a certain $g_1$ range and predicting the $g_1$ value within a different, non-overlapping range for unseen test data.
More precisely, we consider two cases: the training (test) data with~$0<g_1<0.5$ ($0.8<g_1<0.9$) in Case~$1$, and the training (test) data with~$0.5<g_1<1.0$ ($0.1<g_1<0.2$) in Case~$2$. Fig.~\ref{fig:qps_coupling} shows the predicted and true $g_1$ coupling constants, averaged over all experiments, for the two cases. The QCNN model can predict the $g_1$ values with good accuracy, though we observe a small offset for the test data in Case~$1$.
 
Averaging over all the test data, the prediction accuracy is calculated as the absolute difference $\Delta g_1$ between the predicted and true coupling constants. For the QCNN (HEA) with $N_L=3$ ($N'_L = 6$), the prediction accuracies are obtained to be $0.20\pm0.01$ ($0.18\pm0.01$) and $0.07\pm0.01$ ($0.07\pm0.01$) in Case $1$ and $2$, respectively. Increasing $N'_L$ does not improve the accuracy for the HEA. The overall accuracies defined in this way turn out to be similar between the QCNN and HEA. A more detailed comparison of prediction accuracies and their dependence on~$N_L$ and the number of trainable parameters are described in Appendix~\ref{app:qps}. 

\subsubsection{Classification of fermion flavors}
First, we consider two limit situations in which the classification of the fermion flavor is either trivial or impossible. Then, we identify a scenario in which the classification becomes highly nontrivial. 

In the first case, 
we consider the situation in which 
different values of $g_1$ and $g_2$ are assigned and we aim at learning the $\phi$-boson state generated from an initial fermion selected randomly between $f_1$ and $f_2$. However, we can easily realize that, in this case, the flavor defined by the initial conditions is perfectly classifiable. 
Next, we consider 
the other limit scenario in which 
two sets of coupling constants, $G$ and $G'$, are pre-defined and an initial fermion $f_i$ is selected at random.
Then, depending on the randomly chosen initial fermion, the coupling constants are drawn from either of the two sets; for $f_i = f_{1(2)}$, we draw $g_{1(2)}$ from $G$ and $g_{2(1)}$ from $G'$. The $g_{12}$ value is set to 1.
In this specific situation, the shower profile becomes independent of the initial fermion flavor due to the choice of the coupling constants, therefore the classification will always fail. 
However, building on this last scenario, we observe that if the emission scale $\theta_i$ is modified depending on the fermion flavor, the flavor difference is restored and the classification becomes possible.

Moving one step further, one can think of a setup where the emission scale is modified at only one step chosen randomly between 1 and $N_{\rm step}$, for the specific flavor of fermion, e.g., $f_1$. This situation is analogous to the case where the $\phi$ boson is radiated once by a slightly modified mechanism due to an unknown flavor-dependent physics process. In order to detect such processes, one wants to maximize the probability $p$ that the fermion flavor is correctly classified.
If no such modification is made to the emission scale, there is no distinction between radiation processes in terms of fermion flavors, resulting in $p=0$, as mentioned above.
If, for instance, the emission scale is modified once by 20\%, the probability becomes~$p = 0.79\pm0.02$ for the QCNN with~$N_L=3$ and~$p = 0.63\pm0.02$ for the HEA with $N_L'=6$ (see details in Appendix~\ref{app:qps}). The QCNN clearly outperforms HEA in this benchmark case. This is qualitatively understood to be due to different circuit structures for the two ansatzes: a tiny modification to the $\phi$-boson state anywhere in the~$N_{\rm step}$-qubit register is propagated to the measured qubit.
Moreover, the propagation could be more efficient for the QCNN due to its characteristic convolution and pooling structure. More detailed results including dependencies on $N_L$ and the number of trainable parameters are given in Appendix~\ref{app:qps}.

\section{Conclusions and Discussion}
\label{sec:conclusion}

In this work, we focused on quantum machine learning problems with quantum data, and specifically extended the QDL framework to high-energy physics, identifying a set of paradigmatic tasks and promising use-cases.

In particular, we applied quantum convolutional neural networks to quantum data sets generated from quantum simulations of LGT and QFT models. First, we considered the problem of quantum phase recognition in the Schwinger model, and we successfully demonstrated the detection of symmetry-breaking phases near a critical point. Second, we considered the $\mathbb{Z}_2$ gauge theory and classified phases from time-evolved states, corresponding to the deconfinement and confinement of fermionic matter depending on the presence of a background $\mathbb{Z}_2$ gauge field. Our results demonstrated non-trivial learning capabilities and a good generalization behavior for the QCNN method, 
in a scenario where no known simple local order parameter associated with symmetry breaking was available.
Finally, we considered multi-particle states simulated by a phenomenological QFT model of parton showers, from which fermion flavors and coupling constants between fermions and bosons were extracted. The QCNN method performed well in both tasks and, in the setup where some unknown flavor-dependent modification was made to the PS process, it achieved superior performances in flavor classification compared to a more generic hardware-efficient ansatz.

Our studies suggest several future research directions. For example, it would be interesting to understand the connection between specific properties of the input quantum data, the structure of the QCNN ansatz, and its trainability. The former can, for instance, prevent various amounts of entanglement or encode certain physical symmetries such as gauge invariance~\cite{PhysRevResearch.3.043209}, which should probably be reflected in the structure of the learning model. At the same time, it is also well known that the choice of the ansatz and, perhaps less intuitively, of the input states can have a significant impact on the trainability of general QML models~\cite{2021arXiv211014753T,leone2022practical,Larocca2022diagnosingbarren}. 

In terms of applications, it would certainly be important to extend our results to more complex models and tasks that are hard to learn with conventional methods. A primary example in this direction could be an investigation of phases
in non-Abelian and/or higher dimensional gauge theories (see e.g.~\cite{PhysRevA.73.022328,2015PhRvD..91e4506Z,2017PhRvA..95b3604Z,2019PhRvD.100c4518L,PhysRevD.102.094501, 
2023arXiv230402527Z,PhysRevD.101.114502,2022Symm...14..305L,
2021arXiv210712769K,2022RSPTA.38010068W, 
Atas:2021ext,PhysRevD.106.074502,
10.21468/SciPostPhys.6.3.029,2023PhRvD.107e4512F,2023PhRvD.107e4513F,PhysRevD.105.074504,2022arXiv220703473A,PhysRevLett.115.240502,
PhysRevD.101.074512,PhysRevD.103.094501, 
2014AnPhy.351..634M,Haase:2020kaj,2022PhRvD.106k4511C,PRXQuantum.3.020320,PRXQuantum.2.030334,Bauer:2021gek,Grabowska:2022uos}
for quantum simulation of such theories).
Recent results pointed out that classical machine learning algorithms with access to samples from quantum systems can, in certain regimes, successfully and efficiently carry out some of the tasks commonly associated with the QDL paradigm such as predicting properties and phases of many-body quantum systems~\cite{Huang_2022,lewis2023improved}. A comparison of the two approaches, aimed at identifying the proper conditions to achieve a practical quantum advantage, would represent a natural continuation of our line of research.

Another set of open questions concerns the generalization power of QDL models, which is an actively investigated topic in the QML literature~\cite{Banchi_2021,Caro_Generaliz_2022,Caro_OutOfDist_2022}. Besides studying the performance of, e.g., QCNNs in the context of more advanced LGT and QFT models than the ones analyzed in this work, some specific follow-up studies on the QPS algorithm can be considered. Precisely, for this model the $g_{12}$ value is crucial for creating the interference between fermion flavors that is hard to simulate classically. In the numerical studies of Sec.~\ref{sec:qps}, the~$g_{12}$ was kept constant at $0$ or $1$. The fermion flavor classification and the~$g_1$ prediction were attempted with these two different~$g_{12}$ values, and the results were not significantly different. However, the direct prediction of~$g_{12}$ values was also investigated on a test dataset with non-overlapping~$g_{12}$ values in the training dataset, and in this case the accuracy was much worse than for the case of predicting~$g_1$ values in Fig.~\ref{fig:qps_coupling}. Understanding this behavior could provide more insight into the generalization ability of QCNN ansatzes when an intrinsically quantum mechanical property of parton showers is involved.

Finally, a fascinating direction to look at is the exploration of the QDL paradigm in connection with future experiments, where quantum transduction techniques to coherently convert a quantum state from one quantum system to another -- possibly from a detector to a quantum information processing device -- may become available. This would potentially offer exciting opportunities for effectively characterizing quantum phenomena observed in experiments or collected by quantum sensors~\cite{huang2022quantum,aharonov2022quantum}, and, for instance, extracting dynamical properties of interacting particles in high-energy physics. Here, the applications of QDL methods may once more provide a promising path to shed light on complex quantum systems.

\begin{acknowledgments}
We thank Anna Phan for support during the early stages of this project.
This work as well as LN is partly supported by IBM-UTokyo lab.
IBM, the IBM logo, and ibm.com are trademarks of International Business Machines Corp., registered in many jurisdictions worldwide. Other product and service names might be trademarks of IBM or other companies. The current list of IBM trademarks is available at \url{https://www.ibm.com/legal/copytrade}.
\end{acknowledgments}

\appendix
\section{Structures of QCNN circuits}
\label{app:qcnn}

The basic structure of QCNN circuit is common in the studies presented in Sec.~\ref{sec:schwinger}, \ref{sec:z2} and \ref{sec:qps}, each of those referred hereafter to as the Schwinger, $\mathbb{Z}_2$ and QPS, respectively. In general, the QCNN circuit is composed of alternating layers of the convolution (CL) and pooling (PL) unitaries, each constructed with repeated blocks of gates, followed by a fully-connected layer (FCL) and a measurement operator, as in Fig.~\ref{fig:QPS-QCNN-circ}. In the studies presented in this paper, one of the qubits from the last PL block is directly measured, therefore the FCL is not used. Moreover, the gates and the qubit connections to implement each CL and PL blocks are chosen slightly differently to be suited for individual studies.

\subsection{Convolution layer}

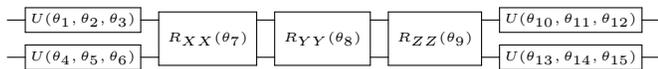
\begin{figure}[htbp]
    \centering
    \tiny
    \Qcircuit @C=1em @R=.7em {
 	& \gate{U(\theta_1,\theta_2,\theta_3)} & \multigate{1}{R_{XX}(\theta_7)} & \multigate{1}{R_{YY}(\theta_8)} & \multigate{1}{R_{ZZ}(\theta_9)} & \gate{U(\theta_{10},\theta_{11},\theta_{12})} & \qw \\
 	& \gate{U(\theta_4,\theta_5,\theta_6)} & \ghost{R_{XX}(\theta_7)} & \ghost{R_{YY}(\theta_7)} & \ghost{R_{ZZ}(\theta_7)} & \gate{U(\theta_{13},\theta_{14},\theta_{15})} & \qw 
    }
    \caption{A general SU(4)-like gate used in a CL block. A $R_{SS}(\theta)$ gate on qubits $i$ and $j$ corresponds to the operator $e^{-i\frac{\theta}{2}\sigma_i^S\sigma_j^S}$ where $\sigma_i^S$ is a Pauli-$S$ ($\in\{X,Y,Z\}$) operator acting on qubit $i$. 
    }
    \label{fig:conv_gate}
\end{figure}

An individual block in the CL consists of a generic SU(4)-like gate with 15 independent parameters. This is common in all the studies and each block is implemented using single- and two-qubit gates with rotation angles as parameters (Fig.~\ref{fig:conv_gate}). 

For the Schwinger and $\mathbb{Z}_2$ studies, the $U$ gate is implemented with a generic single-qubit gate 
$$
U(\theta_1,\theta_2,\theta_3)= 
\begin{pmatrix}
\cos(\theta_1/2) & -e^{i\theta_3}\sin(\theta_1/2) \\
e^{i\theta_2}\sin(\theta_1/2) & e^{i(\theta_2+\theta_3)}\cos(\theta_1/2) \\
\end{pmatrix}.
$$
The two-qubit gates act on neighboring qubits in the first layer, but then on further separated qubits as the CL block goes deeper in the circuit, as shown in Figs.~\ref{fig:qcnn_schwinger_gates} and \ref{fig:qcnn_z2_gates}.
For the QPS study, the $U$ gate is constructed using Pauli operators as $U(\theta_1,\theta_2,\theta_3)=R_Z(\theta_3)R_Y(\theta_2)R_X(\theta_1)$ and the two-qubit gates always act upon neighboring qubits (see Fig.~\ref{fig:qcnn_qps_gates}). The boundary conditions for the CL blocks are different: the Schwinger uses the open boundary condition, while the $\mathbb{Z}_2$ and QPS use the closed one, as indicated by the truncated $U_{\rm conv}$ gates in Figs.~\ref{fig:qcnn_z2_gates} and \ref{fig:qcnn_qps_gates}.

\subsection{Pooling layer}

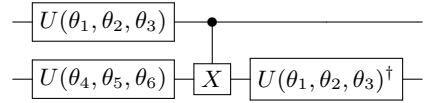
\begin{figure}[htbp]
    \centering
    \centerline{
    \Qcircuit @C=0.8em @R=.7em {
 	& \gate{U(\theta_1,\theta_2,\theta_3)} & \ctrl{1} & \qw & \qw  \\
 	& \gate{U(\theta_4,\theta_5,\theta_6)} & \gate{X} & \gate{U(\theta_1,\theta_2,\theta_3)^\dagger} & \qw 
    }}
    \caption{A set of single-qubit rotation gates and a CNOT gate that composes a PL block. The single-qubit gates on the target qubit after the CNOT correspond to the adjoint of the set of single-qubit gates on the control qubit. 
    }
    \label{fig:pool_gate}
\end{figure}

\begin{figure}[htbp]
    \centering
    \includegraphics[width=0.85\columnwidth]{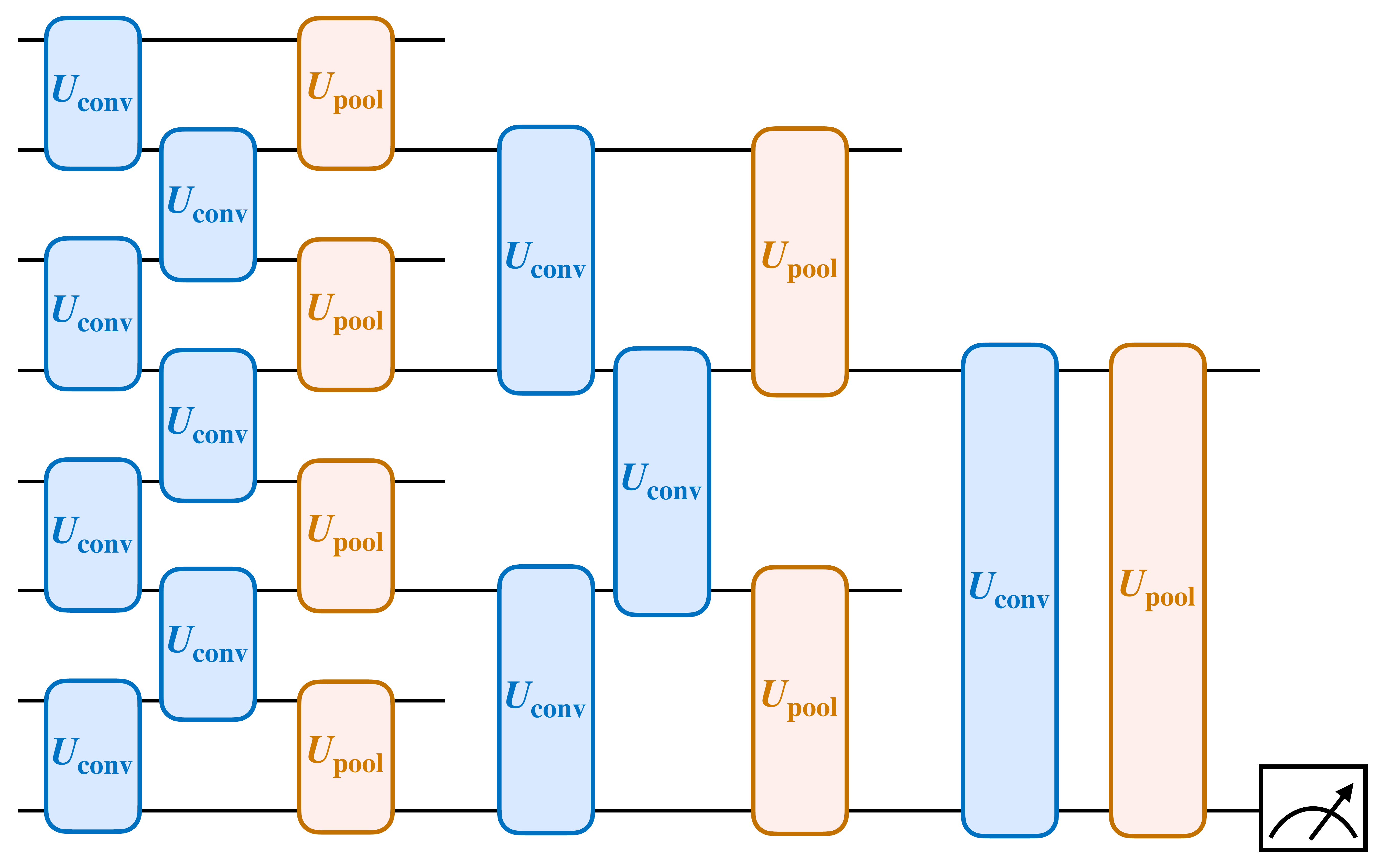}
    \caption{QCNN circuit used for the phase recognition task in the Schwinger model. The unitaries $U_{\rm conv}$ and $U_{\rm pool}$ correspond to those shown in Figs.~\ref{fig:conv_gate} and \ref{fig:pool_gate}, respectively.}
    \label{fig:qcnn_schwinger_gates}
\end{figure}

\begin{figure}[htbp]
    \centering
    \includegraphics[width=0.6\columnwidth]{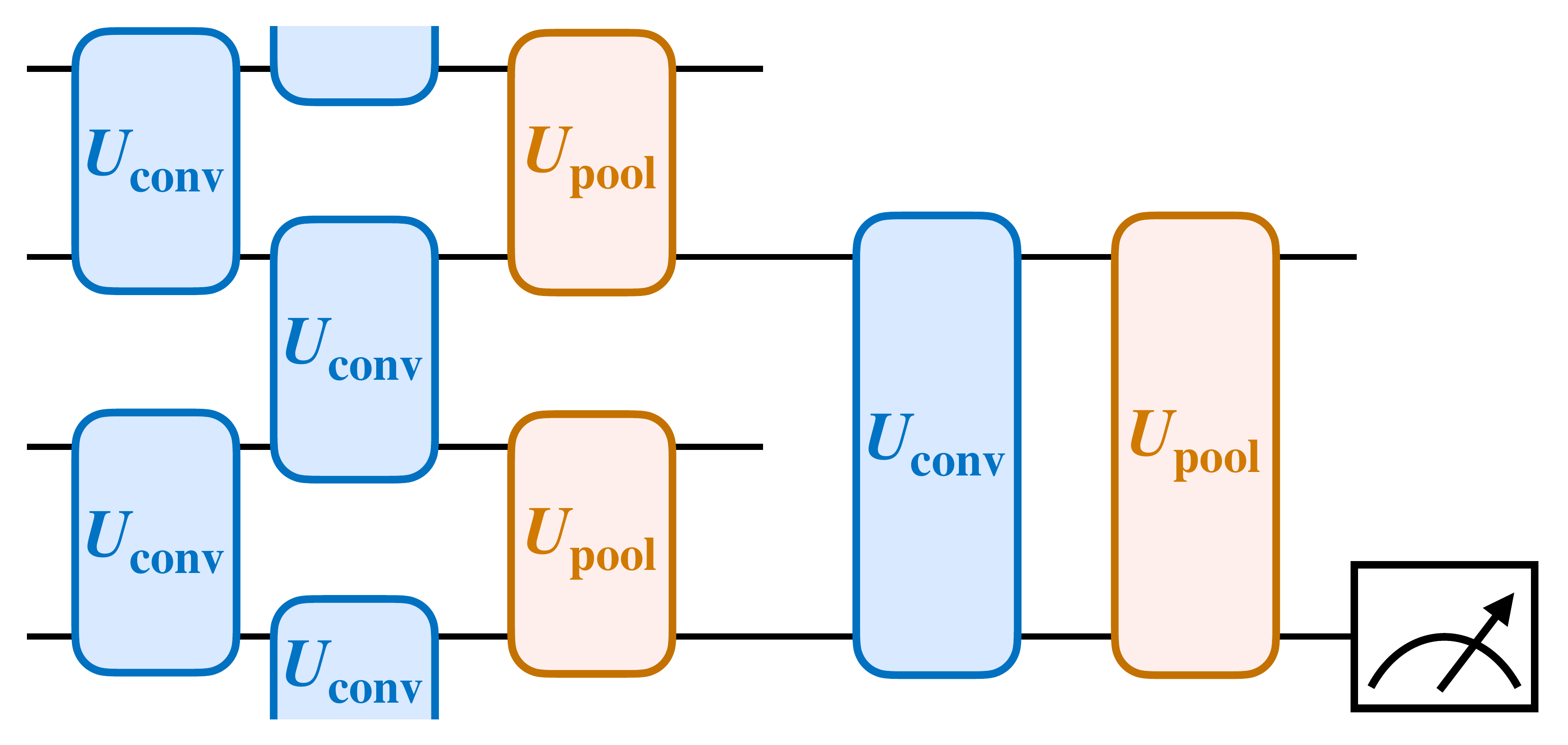}
    \caption{QCNN circuit used for the phase recognition task in the $\mathbb{Z}_{2}$  model. The unitaries $U_{\rm conv}$ and $U_{\rm pool}$ correspond to those shown in Figs.~\ref{fig:conv_gate} and \ref{fig:pool_gate}, respectively. The truncated $U_{\rm conv}$ gate at the first CL indicates that the gate acts upon the top and bottom qubits. }
    \label{fig:qcnn_z2_gates}
\end{figure}

For the PL, each block consists of a set of single-qubit rotation gates acting on the control and target qubits, followed by a CNOT gate and the adjoint of the single-qubit gate set on the target qubit (Fig.~\ref{fig:pool_gate}).
As for the single-qubit gate, the $U$ gates used in the CL blocks above are also used here. The qubit connections for the control and target qubits follow the same rule as those used in the CL blocks for the Schwinger and $\mathbb{Z}_2$. For the QPS, the control and target qubits are separated by $N_{\rm step}/(2L)$ in qubit counts where $L$ is the number of PLs, and the block is repeatedly applied by incrementing the qubit locations by one in a given PL (see Fig.~\ref{fig:qcnn_qps_gates}).

\subsection{QCNN circuit}

\begin{figure}
    \centering
    \includegraphics[width=\columnwidth]{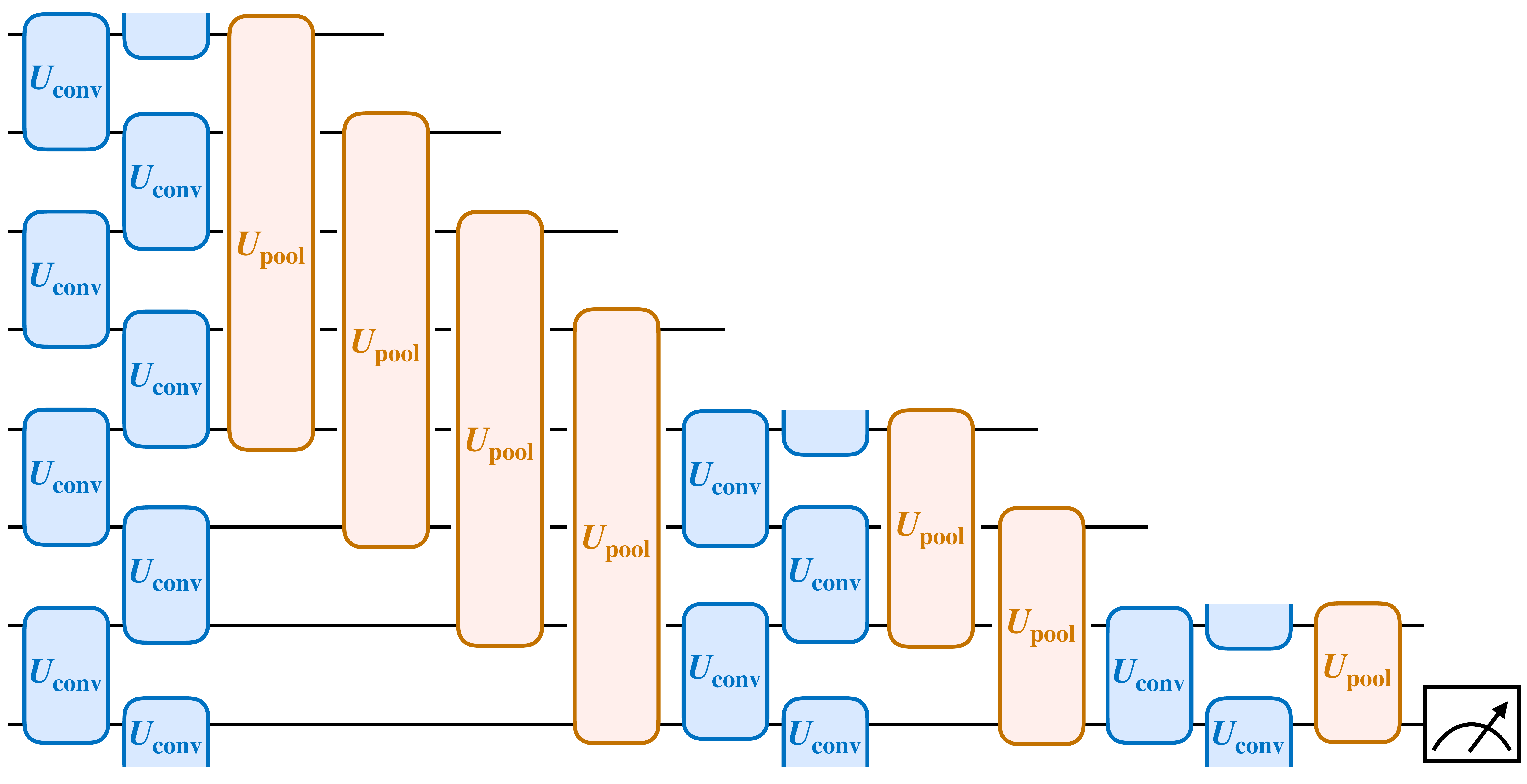}
    \caption{QCNN circuit used in the QPS study with $N_{\rm step}=8$. The unitaries $U_{\rm conv}$ and $U_{\rm pool}$ correspond to those shown in Figs.~\ref{fig:conv_gate} and \ref{fig:pool_gate}, respectively. The truncated $U_{\rm conv}$ gates at the same CL indicate that they act upon those two qubits.}
    \label{fig:qcnn_qps_gates}
\end{figure}

The QCNN circuits used for the Schwinger, $\mathbb{Z}_2$ and QPS studies are shown in Figs.~\ref{fig:qcnn_schwinger_gates}, \ref{fig:qcnn_z2_gates} and \ref{fig:qcnn_qps_gates}, respectively.
For the Schwinger and QPS ($N_{\rm step}=8$) the set of CL and PL is repeated three times, while for the $\mathbb{Z}_2$ it is repeated two times. For all the three cases, the last qubit is always measured. The boundary conditions and the qubit connections to the convolution and pooling blocks are different as in the figures.
Except for the $\mathbb{Z}_2$ gauge theory simulation, the sets of parameters for the CL and PL in the same layer are same, making the circuit translationally invariant.

\section{VQE for the Schwinger model}\label{sec:HVA-schwinger}
This appendix explains the ansatz we used for preparing dataset in the Schwinger model.
We used the so-called Hamiltonian variational ansatz (HVA)~\cite{PhysRevA.92.042303, 10.21468/SciPostPhys.6.3.029, PRXQuantum.1.020319}.
The ansatz is defined by $U_\text{HVA}(\bm{\lambda})\ket{\phi}$ where
\begin{widetext}
\begin{align}
U_{\text{HVA}}(\bm{\lambda})
&=\prod_{l=0}^{N'_L-1}
\Big[\exp\left(
-\ii \lambda_{l}^{(0)}H_{Z}
\right)
\exp\left(
-\ii \lambda_{l}^{(1)}H_{XY}^{(\text{odd})}
\right)
\exp\left(
-\ii \lambda_{l}^{(2)}H_{XY}^{(\text{even})}
\right)
\Big]
\,,
\end{align}
\end{widetext}
with
\begin{align}
 H_{Z} &= \sum_{n=0}^{N_s-2} \left[\sum_{i=0}^{n}\frac{Z_i + (-1)^i}{2}+\frac{\theta}{2\pi}
 \right]^2 
 + \frac{m}{2}\sum_{n=0}^{N_s-1}(-1)^n Z_n\,,
 \end{align}
 \begin{align}
 H_{XY}^{(\text{odd})}
& =
 \sum_{m: \text{odd}}\big[X_{2m-1} X_{2m}+Y_{2m-1}Y_{2m}\big]\,,
 \\
 H_{XY}^{(\text{even})}
& =
 \sum_{m: \text{even}}\big[X_{2m} X_{2m+1}+Y_{2m}Y_{2m+1}\big]\,.
\end{align}
{The initial state was fixed as $\ket{\phi} = \ket{01}^{N_s/2}$ to constrain the minimization to the zero-magnetization sector $\sum_i Z_i \ket{\psi}= 0$}.
\section{Exact diagonalization analysis of the Schwinger model}
\subsection{Critical mass for a finite lattice spacing}
\label{subsec:critical-mass-schwinger}
As mentioned in the main text, the critical mass in the continuum limit is already obtained in~\cite{Hamer:1982mx,Byrnes:2002nv,Dempsey:2022nys}.
Here we want to obtain the critical mass for a finite lattice spacing $a$.
For this purpose, we use the finite scaling analysis~\cite{hamer1988finite} and exact diagonalization (ED).
First, we compute the mass gap $\Delta_N(x)$ with changing the value of mass $x=m/g$, and then obtain the following ratio:
\begin{equation}
    R_N(x)
    =\frac{N\Delta_N(x)}{(N+2)\Delta_{N+2}(x)}\,.
\end{equation}
We choose five points near $R_N=1$ and solve $R_N(x^*_N)=1$ by a linear fitting.
We repeat this process for $N\in\{10,12,14,16,18,20\}$ and extrapolate the results to the limit of $N\to\infty$.
This gives the critical mass for $ag=2$ as $(m/g)\approx0.143$.
\subsection{Local obseravable in the Schwinger model}
\label{subsec:local-observable-schwinger}
\begin{figure}
    \centering
    \includegraphics[width=80mm]{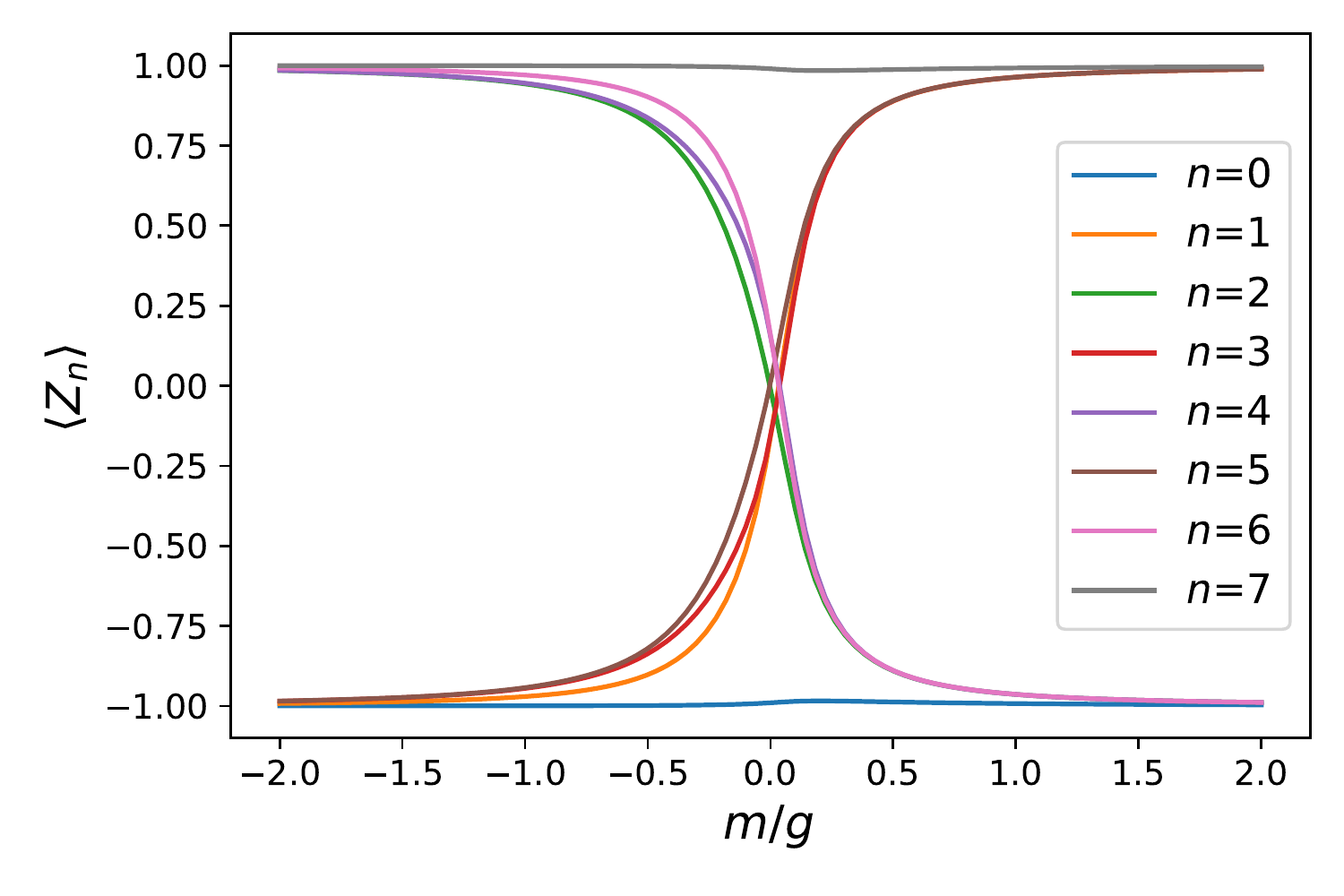}
    \caption{The expectation value of the Pauli operator $\langle Z_n\rangle$ at $n$-th qubit for the Schwinger model with $N_s=8,ag=2,\vartheta=\pi$.}
    \label{fig:vev-of-z-schwinger}
\end{figure}

As mentioned in the main text, the phase transition in the Schwinger model can be detected by measuring the electric field operator.
On top of that, a similar profile is already observed in $\braket{Z_n}$ in the bulk, as shown in Fig.~\ref{fig:vev-of-z-schwinger}.
This prevents us from seeing a clear advantage in terms of sampling complexity because we can roughly distinguish two phases by just a single qubit measurement.
However, it would be worth noting that, for the QCNN model, we measure the last qubit and distinguish two phases, which we cannot do otherwise by a single qubit measurement. In this sense, the QCNN ``sends'' the information in the bulk to the boundary.
\section{Quantum circuit for real-time evolution in $\mathbb{Z}_2$ gauge theory}
\label{sec:circuit-z2}
This appendix explains the detail of dataset preparation in the~$\mathbb{Z}_2$ gauge theory.

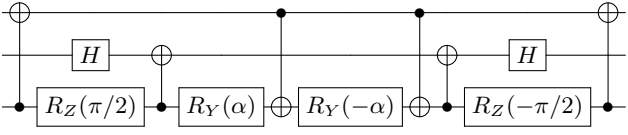
\begin{figure}
    \centering
    \Qcircuit @C=0.3em @R=.7em {
& \targ & \qw & \qw & \qw & \ctrl{2} &\qw & \ctrl{2} & \qw &\qw &\targ &\qw\\
& \qw &  \gate{H} & \targ & \qw & \qw &\qw &\qw& \targ &\gate{H} &\qw &\qw\\
& \ctrl{-2} & \gate{R_{Z}(\pi/2)} & \ctrl{-1} & \gate{R_{Y}(\alpha)} & \targ & \gate{R_{Y}(-\alpha)}  &\targ &\ctrl{-1} &\gate{R_{Z}(-\pi/2)} & \ctrl{-2} &
\qw
}
    \caption{Quantum circuit for $e^{-i(XZX+YZY)\alpha/2}$}
    \label{fig:circuit-z2-trotter}
\end{figure}

First, we decompose the time-evolution operator of $\mathbb{Z}_2$ gauge theory via the Suzuki-Trotter formula as
\begin{equation}
    e^{-iH\Delta t}
    \approx
    e^{-iH_f\Delta t}e^{-iH_g\Delta t}e^{-iH_m\Delta t}\,,
\end{equation}
where 
\begin{align}
H_f=&-\frac{J}{2}\sum_{n=0}^{N_s-1}({X}_{n}{Z}_{n,n+1}{X}_{n+1}+{Y}_{n}{Z}_{n,n+1}{Y}_{n+1})\,,
\\
H_g=&-f\sum_{n=0}^{N_s-1}{X}_{n,n+1}\,,
\\
H_m&=
\frac{m}{2}\sum_{n=0}^{N_s-1}(-)^{n}{Z}_{n}\,.
\end{align}
The latter two factors, $e^{-iH_g\Delta t}$ and $e^{-iH_m\Delta t}$, are easily implemented using single-qubit rotation gates of $R_X$ and $R_Z$, while the first factor $e^{-iH_f\Delta t}$ can be implemented, following~\cite{2004quant.ph..1178V}, by a quantum circuit shown in Fig.~\ref{fig:circuit-z2-trotter}.

Second, the initial state $\ket{\psi_\text{init}}$ which satisfies Gauss's law in Eq.~\eqref{eq:gauss-law-z2} with $g_n=-1$ for $n\neq1$ and $g_n=+1$ for $n=1$ can be obtained by
\begin{equation}
    \ket{\psi_\text{init}} = \prod_{n=0}^{N_s-2}H_{n,n+1}\prod_{n\neq1}^{N_s-1}X_n\ket{0}\,.
\end{equation}

\section{Additional results for QPS simulation}
\label{app:qps}

Here we describe the QPS simulation method used in Sec.~\ref{sec:qps} in more detail.
Fig.~\ref{fig:QPS-QCNN-circ} in the main text shows the QPS circuit used to produce the dataset. The unitary operators $U$ and $U_i^{a/b}$ in the circuit are given by the matrices:
\begin{equation}
U= 
\begin{pmatrix}
\sqrt{1-u^2} & u \\
-u & \sqrt{1-u^2} \\
\end{pmatrix},
\end{equation}
where
\begin{align}
u&=\sqrt{\frac{g_1-g_2+g'}{2g'}},\\
g'&=\text{sign}(g_2-g_1)\sqrt{(g_1-g_2)^2+4g_{12}^2}
\end{align}
and
\begin{equation}
U_i^{a/b}= 
\begin{pmatrix}
\sqrt{\Delta_{a/b}(\theta_i)} & -\sqrt{1-\Delta_{a/b}(\theta_i)} \\
\sqrt{1-\Delta_{a/b}(\theta_i)} & \sqrt{\Delta_{a/b}(\theta_i)} \\
\end{pmatrix},
\end{equation}
where $\Delta_{a/b}(\theta_i)$ represents a Sudakov factor that describes the probability to have no emission from a fermion of type $a/b$ at the $i$-th emission scale $\theta_i$~\cite{Nachman_2021}.
The emission scale is chosen to start from $\theta_{\rm max}=1$ and decrease with increasing number of emission steps, down to the smallest scale of $\epsilon=10^{-3}$.

In addition to the main QCNN circuit, two slightly modified QCNN circuits are prepared for this task. The circuit, denoted as QCNNm1, is formed simply by reducing the number of pairs of the convolution and pooling layers, $N_L$. Another circuit called QCNNm2 is constructed by replacing the gates in the convolution and pooling layers with more simplified ones: the single-qubit gates are reduced to $R_Y$ and $R_Z$ gates in the convolution and pooling layers and the two-qubit gates in the convolution layer to just $ZZ$ gate. The number of trainable parameters $N_{\rm par}$ is $21N_L$ for the QCNN and QCNNm1 and $12N_L$ for the QCNNm2. For the QPS dataset with $N_{\rm step}=8$, $N_L=3$ for the QCNN and 1 or 2 for the QCNNm1. 

\begin{figure*}[htbp]
    \centering
    \includegraphics[width=0.96\columnwidth]{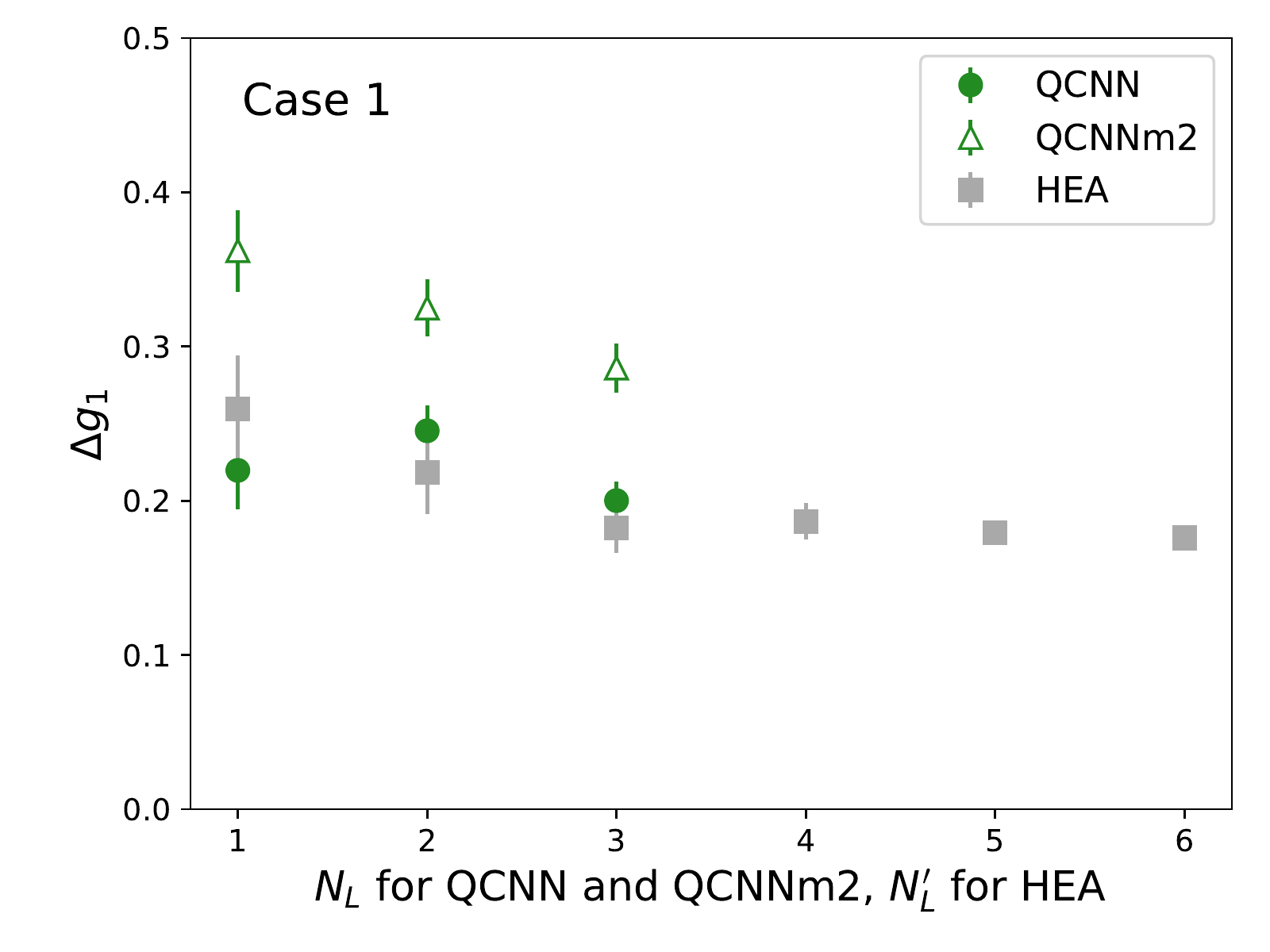}
    \includegraphics[width=0.96\columnwidth]{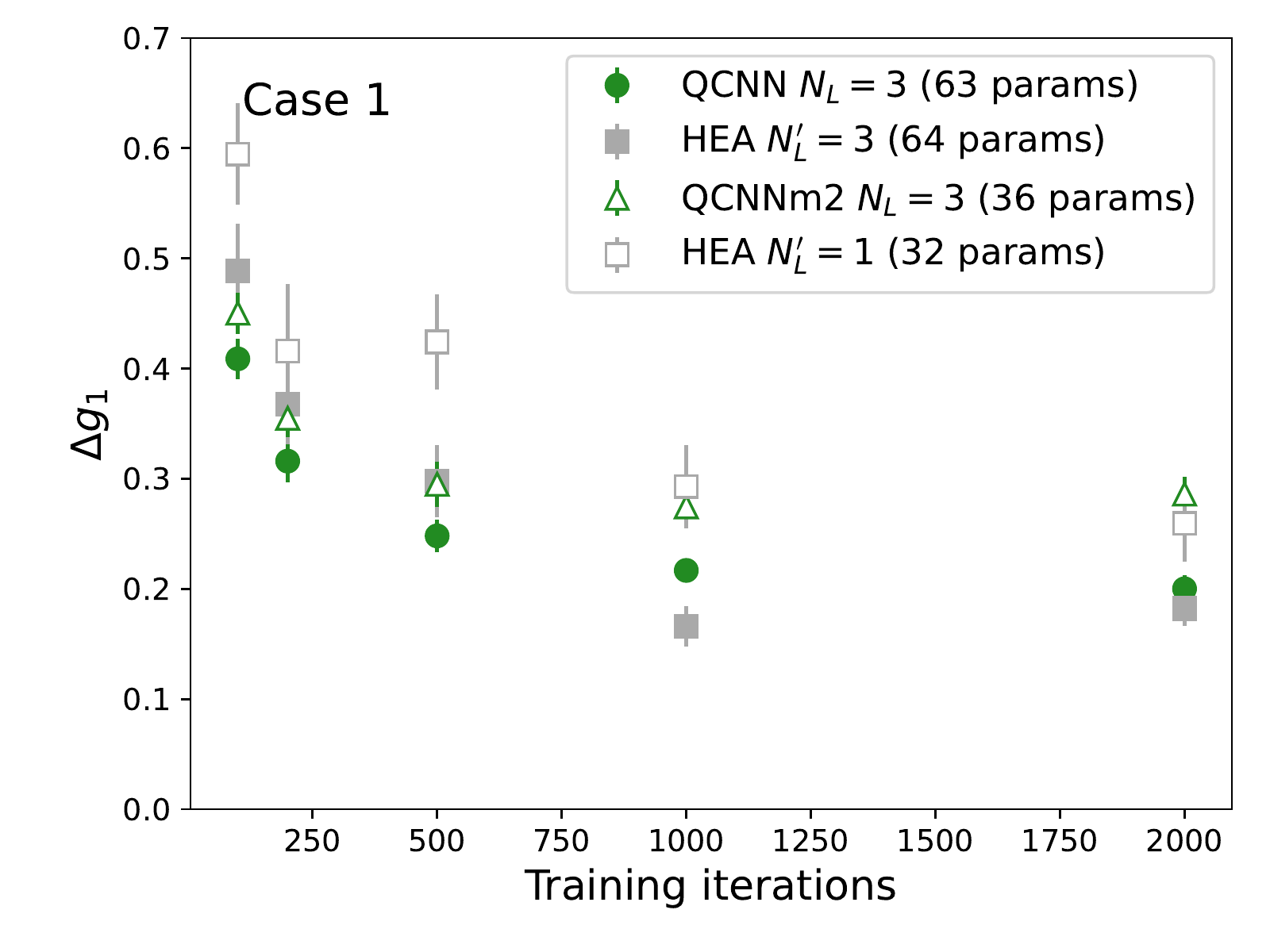}\\
    \includegraphics[width=0.96\columnwidth]{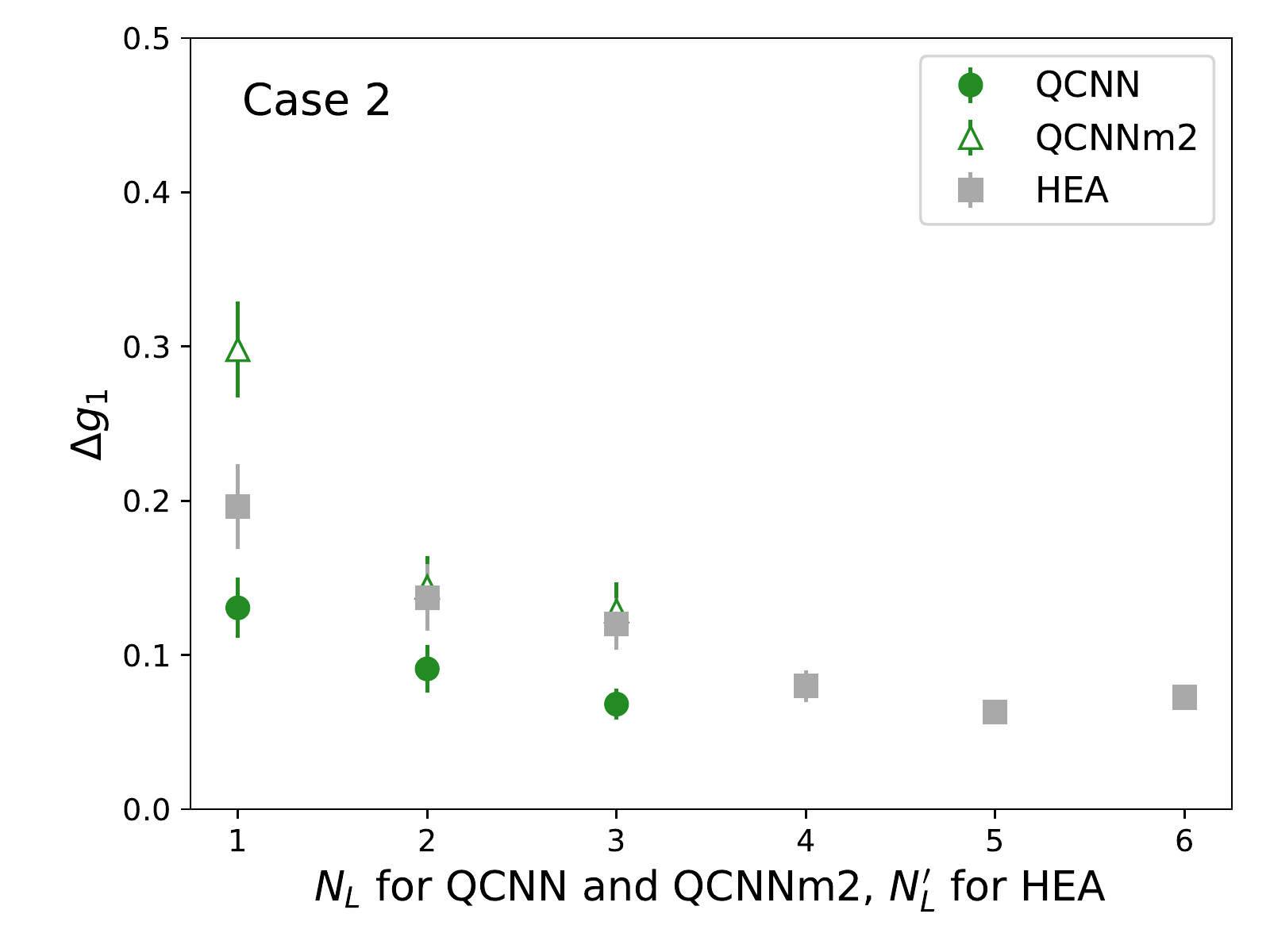}
    \includegraphics[width=0.96\columnwidth]{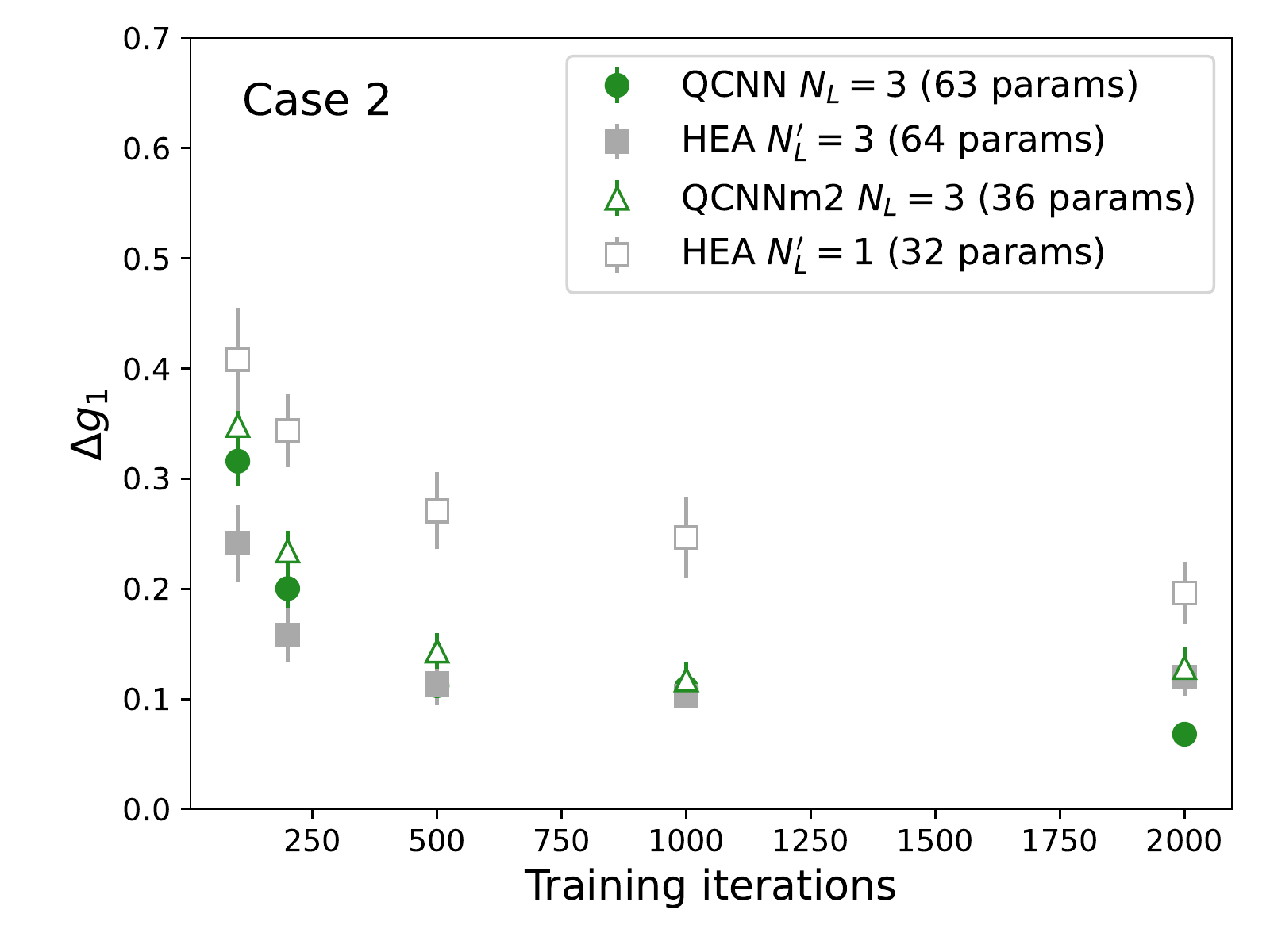}
    \caption{Determination of $g_1$ values at $N_{\rm step}=8$. The absolute difference $\Delta g_1$ between the predicted and true coupling constants is shown for the test data as a function of $N_L$ for QCNN and QCNNm2 and $N_L'$ for HEA, respectively (left) and training iterations (right). The top (bottom) figures show the Case 1 (2) of the training data within $0<g_1<0.5$ ($0.5<g_1<1.0$) and the test data within $0.8<g_1<0.9$  ($0.1<g_1<0.2$). The other coupling constants are set to $g_2=1$ and $g_{12}=0$.}
    \label{fig:qps_coupling_vs_depth}
\end{figure*}

The prediction accuracy, as defined in the main text, is shown as a function of $N_L$ for the QCNN and $N_L'$ for the HEA, respectively, in Fig.~\ref{fig:qps_coupling_vs_depth}. For both selections of the $g_1$ ranges, the prediction accuracy improves with increasing $N_L$ and reaches similar precisions at $N_L=2$ ($N_L'=6$) for the QCNN (HEA). The prediction accuracy and the convergence with the number of training iterations are also compared between the QCNN and HEA models with a similar number of trainable parameters by adjusting the $N_L$ and $N_L'$. The overall behavior of the convergence is similar, but the QCNN model appears to predict more accurately than the HEA when compared at the similar $N_{\rm par}$, in particular at small $N_{\rm par}$.

\subsubsection{Classification of fermion flavors}
\begin{figure*}[htbp]
    \centering
    \includegraphics[width=0.96\columnwidth]{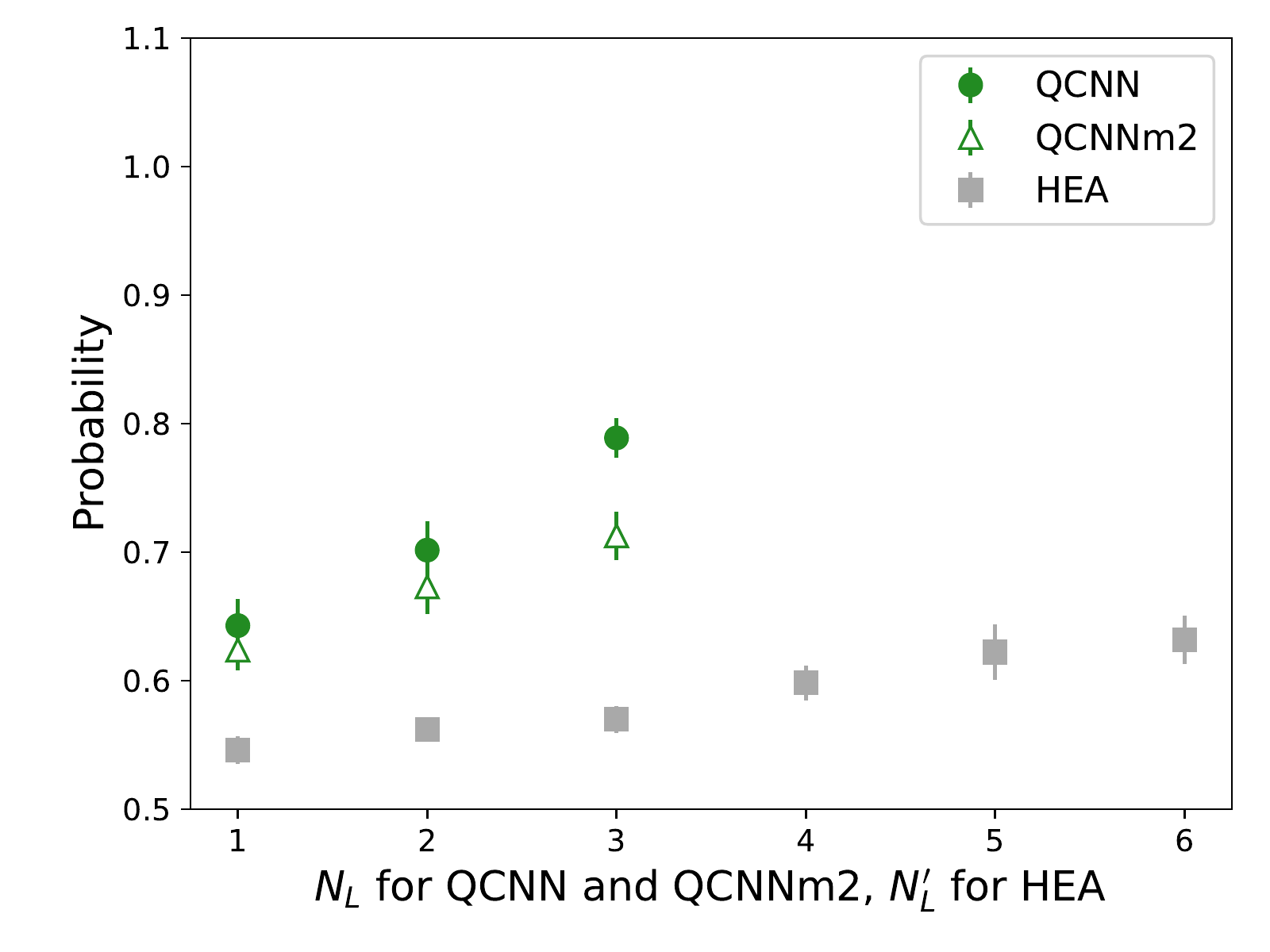}
    \includegraphics[width=0.96\columnwidth]{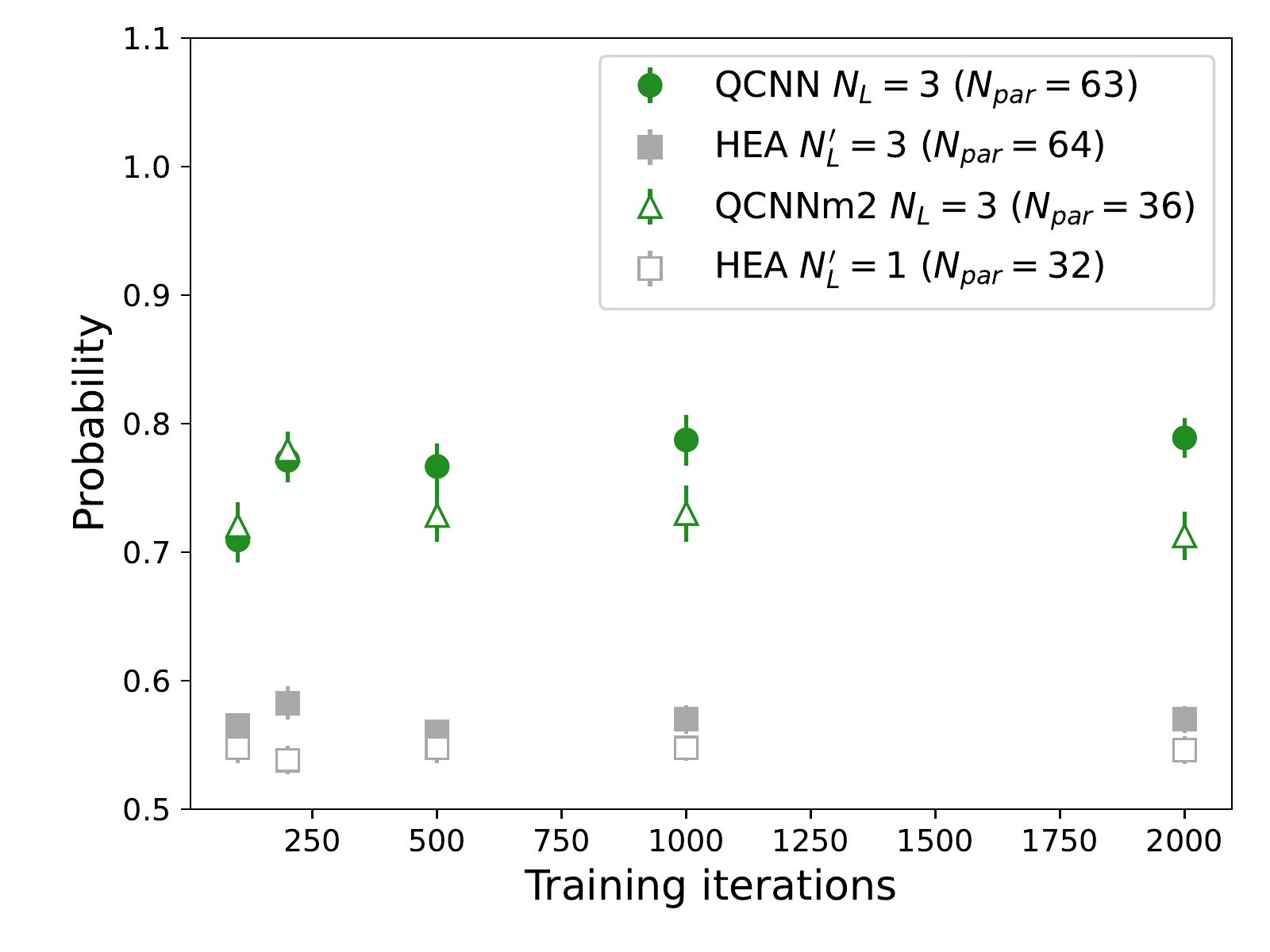}
    \caption{Classification of fermion flavors at $N_{\rm step}=8$. The probability that the fermion flavor is correctly classified when the emission scale is modified only once, randomly between 1 and $N_{\rm step}$ in the shower steps, is shown as a function of $N_L$ for QCNN and QCNNm2 and $N_L'$ for HEA, respectively (left) and training iterations (right).}
    \label{fig:qps_class_vs_depth}
\end{figure*}

Fig.~\ref{fig:qps_class_vs_depth} shows the probability $p$ of correctly classifying the fermion flavors in the scenario that the emission scale is modified only once by 20\%, randomly between 1 and $N_{\rm step}$. The probability increases with $N_L$ for both ansatzes, and the QCNN with $N_L=3$ ($p=0.79\pm0.02$) clearly outperforms the HEA with $N_L'=6$ ($p=0.63\pm0.02$) in this benchmark case. Even the simple QCNNm2 with $N_L=2$ ($p=0.71\pm0.02$) or the QCNNm1 with $N_L=1$ ($p=0.64\pm0.02$) is competive to the HEA. The classification performance versus training iterations is compared between the two ansatzes with similar number of trainable parameters. No significant difference is seen in the convergence of classification performance, but the QCNN can clearly achieve a better performance than the HEA when both ansatzes have similar number of trainable parameters.

\bibliographystyle{utphys}
\bibliography{main}

\end{document}